\documentclass[preprint]{aastex}
\usepackage{graphicx}
\usepackage{threeparttable}
\usepackage{multirow}
\usepackage{hyperref}
\usepackage{color}

\newcommand{\kms}{${\rm km \ s^{-1}}$}
\newcommand{\kmsb}{${\rm km \ s^{-1}}$ }
\newcommand{\nh}{$n_{\rm H}$}
\newcommand{\NH}{$N_{\rm H}$}

\begin{document}

\title{Photoionization-driven Absorption Lines Variability in Balmer Absorption Line Quasar LBQS 1206+1052}

\author{Luming Sun\altaffilmark{1,2}, Hongyan Zhou\altaffilmark{1,2}, Tuo Ji\altaffilmark{1}, Peng Jiang\altaffilmark{1}, Bo Liu\altaffilmark{1,2}, Wenjuan Liu\altaffilmark{3}, Xiang Pan\altaffilmark{1,2}, Xiheng Shi\altaffilmark{1}, Jianguo Wang\altaffilmark{3}, Tinggui Wang\altaffilmark{2}, Chenwei Yang\altaffilmark{2}, Shaohua Zhang\altaffilmark{1}, Lauren Miller\altaffilmark{4}}
\altaffiltext{1}{Polar Research Institute of China, 451 Jinqiao Road, Shanghai, China, lmsun@mail.ustc.edu.cn}
\altaffiltext{2}{Department of Astronomy, University of Science and Technology of China, 96 Jinzhai Road, Hefei, Anhui, China}
\altaffiltext{3}{Yunnan Observatories, Kunming, Yunnan, China}
\altaffiltext{4}{27 Memorial Dr W, Bethlehem, Pennsylvania, USA}

\begin{abstract}
  In this paper we present an analysis of absorption line variability in mini-BAL quasar LBQS 1206+1052.
  The SDSS spectrum demonstrates that the absorption troughs can be divided into two components of blueshift velocities of $\sim$700 \kmsb and $\sim$1400 \kmsb relative to the quasar rest-frame.
  The former component shows rare Balmer absorption, which is an indicator of high density absorbing gas, thus the quasar is worth follow-up spectroscopic observations.
  Our follow-up optical and near-infrared spectra using MMT, YFOSC, TripleSpec and DBSP reveal that the strengths of the absorption lines vary for both of the two components, while the velocities do not change.
  We reproduce all of the spectral data by assuming that only the ionization state of the absorbing gas is variable and that all other physical properties are invariable.
  The variation of ionization is consistent with the variation of optical continuum from the V-band light-curve.
  Additionally, we can not interpret the data by assuming that the variability is due to a movement of the absorbing gas.
  Therefore, our analysis strongly indicates that the absorption line variability in LBQS 1206+1052 is photoionization-driven.
  As shown from photo-ionization simulations, the absorbing gas with blueshift velocity of $\sim$700 \kmsb has a density in the range of $10^9$ to $10^{10}$ cm$^{-3}$ and a distance of $\sim$1 pc, and the gas with blueshift velocity of $\sim$1400 \kmsb has a density of $10^3$ cm$^{-3}$ and a distance of $\sim$1 kpc.
\end{abstract}

\keywords{galaxies: active---galaxies: evolution}

\section{Introduction}

The spectra of the majority of quasars show absorption lines, which can tell us information of the intermediate gas.
The absorption lines can be divided into three classes according to the widths of the absorption lines (e.g., Weymann et al. 1991): broad absorption lines (BAL, $>$2000 \kms), narrow absorption lines (NAL, $<$500 \kms) and mini-BAL (500--2000 \kms).
All BALs and mini-BALs, as well as a part of NALs, are intrinsic absorption lines that are formed by gas associated with the quasar.
They are good diagnostic tools for the surrounding environment of quasar nuclei.
Most intrinsic absorption lines are blueshifted with respect to the corresponding emission lines, implying that the gas is flowing out of the center.
The study of these blueshifted absorption lines can increase our understanding of outflow and quasar feedback, which play a crucial role in the context of galaxy formation and evolution (e.g., Di Matteo et al. 2005).

It has long been known that BAL and mini-BAL systems are variable on time scales from years to months (e.g., Foltz et al. 1987; Smith \& Penston 1988; Barlow et al. 1989).
Monitoring of BAL samples found that more than 60\% of BAL and mini-BAL quasars display variability on time scales of years, and that the fraction of BALs showing variability increases with increasing observing intervals (Capellupo et al. 2011; Filiz Ak et al. 2013).
Almost all BALs vary in depth, with only a minor portion of BALs exhibiting a measurable change in velocity and width (Grier et al. 2016), indicating that the BAL variability is mainly caused by either movement of absorbing gas across our line-of-sight or a change in ionization state.
Studies of individual sources show evidence for both of the two origins, with some supporting the scenario of moving absorbing gas (e.g., Capellupo et al. 2014; He et al. 2014; Muzahid et al. 2016) and others supporting the scenario of varying ionization state (e.g., Grier et al. 2015; Saturni et al. 2016; Wildy et al. 2016).
{ However, which of the two is dominating among all BAL quasars is still under debate.
Capellupo et al. (2013) found variations in only portions of BAL troughs or in lines that are optically thick, suggesting that at least some of changes are caused by clouds moving across our lines of sight.
On the other hand, Filiz Ak et al. (2014) argued that a large fraction of BAL variability was caused by ionization because the variability amplitude of Al III is larger than C IV.}
Wildy et al (2014) found no strong correlation between the variation and continuum luminosity, and then concluded that ionization change is not important.
However, Wang et al. (2015) investigated a large sample of SDSS quasars and showed that the equivalent widths of the lines decrease or increase statistically when the continuum brightens or dims, supporting ionization driven model.
{ Furthermore, He et al. (2017) used a statistical analysis to demonstrate that at least 80\% of the BAL variability is driven by the variation of ionizing continuum.}

The former studies mostly focus on high-ionization BALs (HiBALs) such as C I and Si IV.
About 15\% of optically selected BAL quasars show low-ionization BALs (LoBALs) besides HiBALs, such as Mg II and Al III, { and the fraction may be more in the infrared selected BAL quasars} (e.g., Voit et al. 1993).
{ There is evidence that LoBAL quasars have higher dust reddening (e.g., Zhang et al. 2010) than non-BAL quasars, suggesting they may be at a transition phase from ultraluminous infrared galaxies to unobscured luminous quasars.
Neutral Helium absorption lines He I* $\lambda$10830, 3889, 3189 (transitions from 2p, 3p, 4p from the metastable 2s level in the He I triplet) provide important information on the absorbing media.
The $\lambda$3889 line was noticed early (e.g. Anderson \& Kraft 1969), while the $\lambda$10830 line was first found in AGN by Leighly et al. (2011).}
Liu et al. (2015, hereafter Liu15) demonstrated that He I* absorption could be detected in most Mg II LoBAL quasars if the spectral signal to noise ratio are high enough.
{ This means that He I* absorptions are more common than we thought.
Liu15 and Ji et al. (2015) found that the ionization state of an irradiated medium can be solely determined by the column density of He I* if the medium is thick enough and the ionizing front is well developed.
The variability of LoBAL quasar has not been substantive examined.
Early test (e.g., Vivek et al. 2014) failed to reach a clear conclusion on the dominating cause of LoBAL variability.
We propose that He I* can distinguish the two possible origins of BAL variability}, because it has advantages compared with traditional ions such as Mg II and Al III:
1. He I* multiplets are well separated and do not suffer the blending problem.
2. The optical depths of He I* multiplets cover a large range, such as the optical depth ratio of He I* $\lambda$10830/He I* $\lambda$3889 = 23.1, and thus they are powerful in determining the covering condition of the absorbing gas and the ionic column density at the same time.
{ Example was given in Wildy et al. (2016), and will be given in this work.}

There is a rare population amongst LoBAL quasars which show intrinsic hydrogen Balmer absorption-lines.
{ It was first noticed in Hutching et al. (2002), and so far there are only about ten reported (see Zhang et al. 2015 for a review).}
The origin of these Balmer absorbers is still a mystery, and former researches thought that it comes from high density medium ($n_H>10^6$ cm$^{-3}$).
{ Balmer absorption lines are rare but important because they can provide unique information on the high density gases in the AGN environment, which is difficult to investigate via other approaches.}
For example, we can obtain the density of the gas from strength of Balmer absorption, and hence the distance of the gas to the central ionizing source can be determined using the density and the ionization parameter (e.g., Zhang et al. 2015; Shi et al. 2016a,b).
{ Moreover, the monitoring of Balmer BAL can offer additional information.}
For example, Shi et al. (2016a) reported the Balmer absorption line variability in SDSS J125942.80+121312.6 due to changes in the covering factor, { and hence obtained the transverse velocity of the absorbing gas.
In this work, we will show that monitoring of a Balmer BAL can help to constrain the density of the absorbing gas and then the distance to the quasar nucleus.}

LBQS 1206+1052 was identified during the large bright quasar survey (Hewett et al., 1995) and was classified as a LoBAL quasar (Gibson et al., 2009).
The Balmer-absorption lines in the quasar were identified by Ji et al. (2012, hereafter Ji12) based on SDSS spectrum.
Ji12 divided the BAL troughs into two components.
One component shows an identical profile in Balmer, He I* and Mg II with its centroid blueshifted by $v\approx-726$ \kms.
The other component is detected in He I* and Mg II with $v\approx-1412$ \kms.
Ji12 concluded that Balmer BALs may originate in a partially ionized region via Ly$\alpha$ resonant scattering pumping, and that the gas has a column density $N_H$ in the range of $10^{21}$ to $10^{22}$ cm$^{-2}$ and a density $n_H$ in the range of $10^6$ to $10^8$ cm$^{-3}$ .
The high brightness of LBQS 1206+1052 (SDSS-i band magnitude of 16.50) makes it an excellent candidate for long-term spectroscopic follow-ups which are crucial to absorption line studies.
We made follow-up spectrographic observations using MMT red channel and YFOSC in 2012 to extend the wavelength coverage to better recover the absorption-free spectrum in H$\alpha$ and Mg II regions.
We also took near-infrared (NIR) spectrum using P200 TripleSpec in Feb 2013 to obtain the He I* $\lambda$10830 absorption line.
Intriguingly, the follow-up observations revealed that the absorption lines have a large amplitude variability.
Thus we made further follow-up observations to study the variability.
We analyzed the absorption line variability in LBQS 1206+1052 by recovering the absorption-free spectra and deriving the absorption troughs (Section 3).
We then examined the two possible origins of variability (Section 4).

\section{Observation and Data Reduction}

LBQS 1206+1052 was observed by SDSS 2.5m telescope on 2003-03-24 with an exposure time of 3165 seconds.
The SDSS spectrum was retrieved from the SDSS Data Release 7 database.
The wavelength coverage of SDSS spectrum is 3800--9200 \AA\ and the spectral resolution is 1500--2500.
We obtained follow-up optical spectroscopy of LBQS 1206+1052 with the Red Channel spectrograph of the 6.5 m Multiple Mirror Telescope (MMT) on 2012-03-01, the Yunnan Faint Object Spectrograph and Camera (YFOSC) of the Gaomeigu 2.4m telescope at the Lijiang station of Yunnan Astronomical Observatories (YNAO) on 2012-05-07 and 2016-01-08, and with the Double-Spec (DBSP) spectrograph of the Hale 200-inch telescope at Palomar Observatory on 2014-04-24.
For the MMT observation, we used a grating with 1200 lines/mm which blazed at 1st/7700, and we set the blaze angle to get a wavelength coverage of 8770--9520 \AA.
The exposure time was 500s.
We used a 1.0x180 slit and obtained a spectral resolution of $\lambda / \Delta \lambda \sim3500$ (FWHM) as measured on the night-sky lines.
For the two YFOSC observations, we used the blue-band grism G14, which provided a wavelength coverage from 3500 to 7500 \AA.
For the 2012 observation, we used a slit with a width of 1.8 arcsec.
Two exposures of 40 minutes each were taken.
For the 2016 observation, we used a slit with a width of 2.5 arcsec.
The total exposure time was $\sim$43 minutes.
The median resolutions are $\sim$500 and $\sim$420 for the 2012 and 2016 observations, respectively.
For the DBSP observation, we used a 600/4000 grating for the blue side, and a 600/10000 grating for the red side, and a D68 dichroic was selected.
This setting yielded a wavelength coverage of 2977--6027 \AA\ and 7821--11181 \AA.
We used a 1.5 arcsec slit during the night.
The median resolution is $\sim$1200 for the blue side and $\sim$2800 for the red side.
The MMT, YFOSC and DBSP spectroscopic data was reduced following the IRAF standard routine.
Wavelength calibration was carried out using He-Ne-Ar lamp for YFOSC and Fe-Ar and He-Ne-Ar lamp for DBSP, which were taken on the same night during the observations.
We used the night sky lines for the wavelength calibration of the MMT Red Channel.
The standard stars were observed for flux calibrations just before or behind the observations of LBQS 1206+1052 each night.

We also took NIR spectra using the TripleSpec spectrograph of the Hale 200-inch telescope on 2013-02-23, 2015-05-26 and 2015-12-28.
The observations were carried out in A-B-B-A dithering mode and the total exposure time was 12, 20 and 9 minutes for the three observations, respectively.
For all of the three observations, slits with a width of 1.0 arcsec were used, and the spectral resolutions were $\sim$2200 in H band, where the He I* $\lambda$10830 line is located.
The data was reduced with the specX package.
The information for all of the optical and NIR spectrometries of LBQS 1206+1052 is listed in Table 1, and the spectra are plotted in Figure 1 in rest-frame according to $z=0.3953$.

LBQS 1206+1052 was observed with HST/COS in 2010-05-08 for an exposure time of 4840 seconds using the G130M grating.
The reduced data was downloaded from the Mikulski Archive for Space Telescopes (MAST) and the four exposures were co-added to produce the final extracted one-dimensional spectra.
The spectrum covers a wavelength range of 1143--1442 \AA\, in which a small range of 1288--1300 \AA\ is not available due to a gap between the two CCD chips.
The spectral resolution of this setting is 16000--21000.

We collected the photometric data to generate the light-curve of this object.
LBQS 1206+1052 was observed by the Catalina Sky Survey (Drake et al. 2009).
We took the data from the second data release of Catalina Survey\footnote{http://nesssi.cacr.caltech.edu/DataRelease/}, which supplies the photometric monitoring data in $V_{CSS}$ band during 2005 and 2013.
We converted the magnitude in $V_{CSS}$ band to Bessel-V band to directly compare them with other observations using the equation:
$V = V_{CSS} + 0.91\times(V-R)^2 + 0.04 = V_{CSS}+0.13$ ($\sigma=0.056$),
where $V-R$ was obtained from synthetic V- and R-band magnitudes using SDSS spectra.
The SDSS photometry in ugriz bands was taken in 2002 with an exposure time of 1 minute each, which was also converted to Bessel-V band according to Jester al. (2005).
We made a follow up observation using the 30 cm aperture Bright Star Survey Telescope (BSST) on 2015-04-23.
We took five exposures with the Bessel-V filter with a total exposure time of 15 minutes.
We combined the five V-band images and did aperture photometry for LBQS 1206+1052, and then made calibration using a nearby standard star.
The light-curve from the photometric data is plotted in Figure 2.
We also plotted the synthetic magnitudes in Bessel-V band using SDSS, YFOSC and DBSP spectra.

\section{Variability of Quasar Absorption Lines}

We compared the observed spectra ($f_{\rm obs}$) in regions around major absorption lines, including Mg II $\lambda\lambda$ 2796, 2803, He I* $\lambda$ 2946, 3189, 3889, 10830 and Balmer lines (H$\alpha$, H$\beta$ and H$\gamma$) in Figure 3.
The variability is notable even without any normalization.
The red part of the He I* $\lambda$10830 absorption trough became deeper in 2015 relative to 2013.
The Mg II absorption trough in YFOSC 2012 spectrum is shallower than in the SDSS 2003 and the DBSP 2014 spectra, indicating a process of becoming weak and then strong.
The variability of Mg II cannot be interpreted as a result of lower spectral resolution of YFSOC spectrum, because the disparity is still significant after we blurred the SDSS and DBSP spectra so that the three have the same resolution.

{ For a further investigation of the absorption lines, we analyzed the spectra via common procedure, including recovering the absorption-free spectra, decomposing emission components, and generating normalized absorption spectra.}

\subsection{Recover the absorption-free spectra using pair-matching method}

We needed to recover the absorption-free spectra ($f_{\rm AbsFree}$) to analyze the variability in more detail.
A reliable method to obtain the absorption-free spectra is the pair-matching method (e.g., Leighly et al. 2011).
This method is based on the similarity of continua and emission line profiles between quasars with and without absorptions.
It is always possible to find enough unabsorbed quasars whose spectra resemble the spectral features surrounding the absorption line of a given absorbed quasar.
This method has a big benefit such that the systematic error of the model can be taken into account.
Liu15 developed the pair-matching method to analyze Mg II and He I $\lambda$3889 BALs.
Using a Monte Carlo simulation, Liu15 demonstrated that the method could precisely measure the BAL parameters like depth and EW and robustly estimate the uncertainty of the parameters.
We applied the pair-matching method to LBQS 1206+1052.
For the Mg II doublet and the He I* optical absorption lines, we basically followed Liu15.
We further developed the method by applying it to Balmer and He I* $\lambda$10830 absorption lines.
The details of the pair-matching method that we used will be described in the appendix, and the main idea is summarized as follows.
We divided four ranges in the spectra: 2500--3300\AA\ for Mg II doublet and He I $\lambda$$\lambda$2946, 3189 lines; 3800--4000\AA\ for He I $\lambda$3889; 4825--4892\AA\ and 6510--6610\AA\ for H$\beta$ and H$\alpha$; 10550--11150\AA\ for He I $\lambda$10830.
For each range, the following things were done:
\\1. We generated a library of unabsorbed quasar spectra;
\\2. We looped over the library to fit the spectra of LBQS 1206+1052 from all the observations simultaneously, and the Gaussian blurred the library spectra with different widths for different observations to account for their spectral resolutions.
Windows affected by absorption were masked during the fitting;
\\3. We selected the ten best fits according to $\chi^2$ as the final absorption-free templates.

The pair-matching results are shown in Figure 4.
The median spectra of the absorption-free templates are plotted in red lines.
We also illuminated the systematic error of this pair-matching method by plotting the ranges (green lines) within 1-$\sigma$ standard deviation of the ten best fits for each data point.
In the following part of the paper, when referring to absorption-free spectrum, we mean the median spectrum of the absorption-free templates.

{ In the previous works using pair-matching method, they calculated the BAL EW and its uncertainty by considering both the random error ($\sigma_{\rm ran}$) and the systematical error ($\sigma_{\rm sys}$).
For each accepted pair-matching model, an EW$_i$ can be calculated.
The mean of all EW$_i$ is a good estimate of the true value of EW, and the standard deviation indicates the systematic error of this pair-matching method.
Then the total measurement of BAL EW can be expressed as $\sigma_{\rm total} = \sqrt{\sigma_{\rm sys}^2 + \sigma_{\rm ran}^2}$.
In this work, we need not only to measure the BAL EW, but also to fit the BAL parameters, such as centroids, widths and integrated optical depths.
We proposed a new method to estimate the uncertainty of BAL parameters in the fitting based on Monte Carlo algorithm.
As the previous method, we also considered both the two origins of errors.}
To account for the random error, we generated fake spectra $f_{\rm fake}$ using observed flux and error as:
\begin{equation}
\rm{f_{\rm fake}(\lambda) = f_{\rm obs}(\lambda) + R(\lambda) \times error_{\rm obs}(\lambda)}
\end{equation}
where R($\lambda$) is a series of random number in standard normal distribution.
To account for the systematic error, we randomly selected one template from the accepted fits for each pair-matching region.
The fitting procedure was then done using the fake spectrum and the absorption-free templates by minimizing $\chi^2$, which is expressed as:
\begin{equation}
\chi^2 = (f_{\rm obs}(\lambda) - f_{\rm model}(\lambda))^2 \times {\rm Weight(\lambda)}
\end{equation}
The computation of Weight($\lambda$) considers both the two origin of errors as:
\begin{equation}
{\rm Weight}(\lambda) = \frac{1}{ {\rm error}^2_{\rm obs}(\lambda) + {\rm Stddev}(\lambda) }
\end{equation}
where Stddev($\lambda$) is the standard deviation spectrum of the accepted absorption-free templates for each pixel.
We repeated the measurement 500 times to derive distributions of the BAL parameters.
{ We referred to the median value of the distribution as ``the best value'' of the BAL parameters, and the interval between 5\% and 95\% ranked values as ``90\% confidence interval''.}

{ We examined whether this method can produce good estimations of BAL parameters and their uncertainties by generating fake absorbed spectra using unabsorbed spectra and given profiles, and then fitting them using the above method, and then comparing the best-fitting BAL parameters with the given ones.
The details are described in the Appendix.
From the testing result, we can say that the 90\% confidence intervals of BAL parameters obtained from our method are reliable.
Throughout this paper, we measured the BAL parameters using this method.}

{ We can measure the BAL EWs following the previous works, and we can also do that using our method.
In the Appendix we tested our method and found that the 90\% confidence interval of BAL EW is also reliable.
This means that when measuring the BAL EWs, our method is as good as the previous method, and thus we also used our method.
In Section 3.3, we attempted to measure the EW ratio between two observations of the same object.
We also used our method though we did not test its behavior in this situation.}

\subsection{Emission components decomposition}

Quasar { UV-optical} spectra typically consist of a power-law components representing thermal emission from the accretion disk, BELs, NELs, a Balmer continuum component for the blue bump in the wavelength region bluer than 3645 \AA, and blended Fe II emission lines.
We built a model in which the emission of LBQS 1206+1052 comes from three regions: the accretion disk, the broad emission line region (BELR) and the narrow emission line region (NELR), and the emission components are expressed as $f_{\rm AD}$, $f_{\rm BELR}$ and $f_{\rm NELR}$ hereafter.
In the model, the Balmer continuum and Fe II emissions are believed to have come from the BELR.
We modeled the optical spectra of SDSS 2003, YFOSC 2012, DBSP 2014 and YFOSC 2016 by a procedure detailedly described in Ji12.
We first fit the emission-line-free region (2855--3010, 3625--3645, 4170--4260, 4430--4770, 5080--5550, 6050--6200 \AA) { using an ``pseudo-continuum'' model consists of} a power-law component, a Balmer continuum component, a blended high-order Balmer emission line component, and a Fe II component, by minimization of $\chi^2$.
The MMT 2012 spectrum does not cover emission-line-free regions, thus we rescaled the DBSP spectrum to fit the MMT spectrum in the ranges of 6300--6450 and 6700--6820 \AA, and derived the pseudo-continuum model for MMT using the rescaled DBSP model.
After subtracting the pseudo-continuum model, we applied a joint fit of the five continuum-subtracted spectra using Gaussians simultaneously by assuming the following:
1. All of the profiles of BELs and NELs are double Gaussian.
2. The profiles of all the BELs in one spectrum are identical.
3. The profiles of each NEL in all the five spectra are identical.
4. The profiles of NELs from the same ion are identical.
5. The flux ratios of [N II], [O III], [Ne III] and [Ne V] doublets are fixed at theoretical values.
In the windows affected by absorption lines, we used the absorption-free spectra in the fitting instead of the observed spectra.
The decomposition results of the optical spectra are shown in Figure 5.

For NIR spectra, { we only concerned region around He I* $\lambda$18380 line.
The spectra here only contain continua, BELs and NELs.}
We fit the continua { phenomenological using power-laws} in the emission-line-free windows of 10200--10500, 11500--12400 and 14200--15000\AA.
The continuum-subtracted spectra around He I $\lambda$10830 were then fitted by using a model consisting of He I $\lambda$10830 BEL, Pa$\gamma$ BEL and He I $\lambda$10830 NEL and by assuming the profile of each is double-Gaussian.
The decomposition results of the NIR spectra are also shown in Figure 5.

\subsection{Normalized absorption spectra}

In Figure 6, we plotted the normalized absorption spectra $f_{\rm abs}$.
{ For the optical spectra}, $f_{\rm abs}$ were normalized by assuming that { the absorber only covers the accretion disk and does not cover any BELR and NELR emissions.}
That is:
\begin{equation}
f_{\rm abs}=1+\frac{f_{\rm obs}-f_{\rm AbsFree}}{f_{\rm AD}}
\end{equation}
{ For the NIR spectra, we temporarily treated the power-law component as emission from accretion disk and also used this equation.}

It can be clearly seen that the deepest position in the He I* $\lambda$10830 absorption trough is at a velocity of $v\sim-1400$ \kmsb relative to quasar rest-frame for all the three observations\footnote{By ignoring the pixels affected by a strong sky line}.
The feature at this velocity are also seen in Mg II, as the deepest two positions in Mg II trough in SDSS and DBSP spectra correspond to positions at $v\sim-1400$ \kmsb for Mg II doublet.
The H$\alpha$ is different because the deepest position is at $v\sim-700$ \kmsb and absorption at $v\sim-1400$ \kmsb is not detected.
We divided the absorption troughs into two components due to the difference between Balmer and other lines, and referred to the two components as the V1400 and V700 components for short according to the blueshift velocity.
It can be clearly seen that both of the components contributes to He I* $\lambda$10830 and $\lambda$3889 absorption troughs, while the Mg II trough is more complicated due to blending of the doublet.
Ji12 decomposed the Mg II trough and found that it also consists of the two components, and the V1400 component of Mg II $\lambda$2803 is blended with the V700 component of Mg II $\lambda$2796.
Thus we concluded that the V700 component is in Balmer, He I* and Mg II absorption lines, and the V1400 component is only in He I* and Mg II.

The absorption troughs in the SDSS 2003 spectrum and in follow-up spectra with time intervals of a decade show similar structures, indicating that the two components remain for over ten years.
By directly comparing the absorption troughs from different observations, we concluded that the velocities of the two components do not vary by time.
On the other hand, the variability in strength is notable after normalization.
The V700 component of He I* $\lambda$10830 clearly became weak from 2013 to 2015, while the V1400 component changed little.
Although the optical spectra suffer from different spectral resolutions, we can quantify the variations in depths by using EWs because EW is less affected by resolution.
We calculated the EW ratios relative to EW in the SDSS spectrum, which shows the deepest absorption, and the results are listed in Table 2.
For the V700 component, the variances in H$\alpha$, H$\beta$ and He I* $\lambda$3889 (integrated in $-1000<v<0$) show the same tendency---it became weak from 2003 to 2012, and recovered in 2014 and 2016.
For the V1400 component, we did not detect variance in He I* $\lambda$10830.
The EW of Mg II also decreased from 2003 to 2012 and increased in 2014, suggesting the Mg II of the V1400 component also has the same variability pattern as the He I* and Balmer of the V700 component, considering that the Mg II trough is dominated by the V1400 component.

\section{The variability: change in ionization state or movement of the absorber?}

We showed that the velocity of both of the two components did not change significantly during a decade, while the strengths of all of the absorption lines varied.
This type of variability may result from either a change in the ionization state or the movement of the absorber.
The first can cause a variation in ionic column density, and the second can cause a variation in the covering factor of the absorbing gas.
We attempted to distinguish the two possible origins by reproducing the data using two models: one assuming that only the ionization state of the absorbing gas is variable and that all of the other physical properties are invariable, and the other assuming that only the covering factor of the absorbing gas is variable.

\subsection{Variable Ionization State Model}

If the absorption line variability is due to variation in ionization state, the variance in strength of the absorption lines should be correlated with variance in continuum flux.
We labeled the observation time of the 8 spectral observations in the V-band lightcurve in Figure 2.
We found it is likely that the continuum flux during SDSS 200303 observation is the lowest amongst the 8 observations according to the Catalina light curve and the synthetic magnitude of the SDSS spectrum.
Although not an imaging result, the synthetic magnitude is reliable because the flux calibration of the SDSS data release 7 has good quality, and because the flux is in agreement with the brightening tendency between SDSS photometry in 2002 and Catalina photometry in 2004.
At the same time, all of the absorption lines are the strongest in the SDSS 2003 spectrum.
The highest flux recorded by the Catalina light curve was in 2013, corresponding to TSpec 2013 spectrum, in which the V700 component of He I* $\lambda$10830 is indeed the weakest among the three TSpec spectra.
This fact also reveals that the optical depth of He I* $\lambda$10830 was not large ($<1$) in 2013.
Thus the corresponding optical depth of He I* $\lambda$3889 line (23 times less) should be very small.
The clear detection of He I* $\lambda$3889 in the SDSS, DBSP and YFOSC 2016 spectrum implies that the V700 component of He I* in 2013 is the weakest amongst all of the observations.
Thus, there is an anti-correlation of the continuum fluxes and absorption depths in 2003 and 2013.
We further checked this anti-correlation.
The variation in EW of He I*, Balmer and Mg II show the same pattern.
They became weaker from 2003 to 2012, and became stronger from 2012 to 2014.
On the other hand, the strengthening in continuum from 2003 to 2012 can be clearly seen in the light curve.
If we adopted the magnitude value from synthetics of DBSP spectrum, the dimming in continuum can also be found from 2012 to 2014.
Though it is not a photometric result, The synthetic magnitude of DBSP 2014 spectrum has good quality due to two facts, one is that we observe the standard star just before the observation of LBQS 1206+1052 with close air mass, the other is that the magnitude agrees with the decreasing tendency in 2013 in CSS light curve.
In summary, we found an anti-correlation in EW of absorption lines and the continuum flux, supporting the variable ionization state model.
For a further examination of this model, we will first model the absorbing troughs and obtain the ionic column densities, and then check if the ionic column densities agree with the photo-ionization model using Cloudy simulations.

\subsubsection{Modeling the absorbing troughs}

The covering conditions of absorbing gases to the central continuum source are essential for measurement of ionic column densities.
As can be seen in Figure 6, the V1400 component of the He I $\lambda$10830 absorption line has negative flux if normalized using the continuum, indicating that the V1400 absorbing gas covers not only the whole accretion disk but also a part of the BELR.
By normalizing using both the continuum and BELs, we concluded that the V1400 absorbing gas covers at least 60 \% of the BELR, and hence the size is comparable to, or larger than, the BELR.
On the other hand, the V700 absorbing gas does not cover such a large fraction of the BELR as the V1400 component does.
If it indeed covered 60\% or more of the BELR, the H$\alpha$ optical depth at the deepest position in SDSS spectrum would be $<$0.45, while the H$\beta$ optical depth would be $>$0.1.
These two are inconsistent considering the theoretical ratio of H$\alpha$/H$\beta$ = 7.26.
The normalized flux of H$\alpha$ and He I* $\lambda$10830 at the velocity of $\sim$700 \kmsb can both reach $\sim$0.1, indicating that the V700 absorbing gas covers at least 90\% of the accretion disk.
Assuming that it covers the whole accretion disk and does not cover BELR at all, we computed the optical depth ratio of H$\alpha$ to H$\beta$ to be $7.0\pm2.1$, which is consistent with the theoretical value.
Thus it is likely that the size of the V700 absorbing gas is comparable to the accretion disk and much smaller than the BELR, and hence the two absorbing gases are different in size.
Considering that the V700 component has Balmer absorptions, which are not seen in the V1400 component, the physical conditions of the two absorbing gases are also different.
Thus it is likely that the two absorption gases are physically independent.

We fit the spectral data in absorption regions by assuming the following:
\\1. For each absorption line in each spectrum, the velocity profile of optical depth is Gaussian for both of the V1400 and V700 components.
\\2. For all of the spectral lines which belong to the same ion for each component, the velocity profiles of optical depths are identical and do not vary by time.
\\3. The ionic column densities are computed from the optical depth profile by:
\begin{equation}
N_{ion}=\frac{m_e c}{\pi e^2 f \lambda} \int \tau_v \rm{d}v
\end{equation}
where $f$ and $\lambda$ are the oscillator strength and the wavelength of each spectral line.
For ions with two or more lines in the same spectrum, the column densities are fixed.
\\4. The covering conditions are the same for different ions, and do not change by velocity or vary by time, while they are different for the two components.
Thus for each absorption line we have:
\begin{equation}
f_{\rm obs} - f_{\rm AbsFree} = f_{\rm AD} \times (1 - f_{\rm resi}^{\rm AD}) + f_{\rm BELR} \times (1 - f_{\rm resi}^{\rm BELR})
\end{equation}
where $f_{\rm resi}^{\rm AD}$ and $f_{\rm resi}^{\rm BELR}$ are the residual fraction after absorption for the two emission sources, which can be expressed as:
\begin{equation}
\left\{
\begin{array}{lr}
 {f_{\rm resi}^{\rm AD} = (1 - {\rm CF_{V700}^{AD} + CF_{V700}^{AD}} \times e^{-\tau_{\rm V700}}) \times
        (1 - {\rm CF_{V1400}^{AD} + CF_{V1400}^{AD}} \times e^{-\tau_{\rm V1400}})} \\
 {f_{\rm resi}^{\rm BELR} = (1 - {\rm CF_{V700}^{BELR} + CF_{V700}^{BELR}} \times e^{-\tau_{\rm V700}}) \times
        (1 - {\rm CF_{V1400}^{BELR} + CF_{V1400}^{BELR}} \times e^{-\tau_{\rm V1400}})}
\end{array}
\right.
\end{equation}
where $\rm{CF_{V700}^{AD}}$, $\rm{CF_{V1400}^{AD}}$, $\rm{CF_{V700}^{BELR}}$ and $\rm{CF_{V1400}^{BELR}}$ are the covering factors of the V700 and V1400 absorbing gases to the accretion disk and BELR, and $\tau_{\rm V700}$ and $\tau_{\rm V1400}$ are the optical depths of the two absorbing gases.
We did not consider the situation that the absorbing gas covers the NELR because the NELR is typically too large to be covered by an absorbing cloud.
The V1400 component must cover the whole accretion disk, thus $\rm{CF_{V1400}^{AD}}=1$.
The situation for V700 component is more complicated: it may only cover part of the accretion disk ($\rm{CF_{V700}^{BELR}}=0$), or cover the whole accretion disk and a small part of the BELR ($\rm{CF_{V700}^{AD}}=1$).

Our model has total 46 free parameters, 23 for each absorbing component: line centers and line widths for three ions (Mg II, He I*, and H(n=2)), a total 16 ionic column densities in 8 spectra, and one covering factor.
For the V700 component, the only free covering factor is $\rm{CF_{V700}^{AD}}$ for the first covering situation and $\rm{CF_{V700}^{BELR}}$ for the second covering situation.
We simultaneously fit the 8 spectra in absorption windows ($-2000$ to $0$ \kmsb for each line) using the method described in Section 3.1.
The spectral resolutions are different amongst different observations, and amongst different wavelengths during the same observation, which are shown in Table 1.
We considered the resolution effect by convolving Gaussians to the models.
We calculated the probability distributions of all the parameters, and listed the best values and 90\% confidence intervals in Table 4.

We first set all the parameters free.
Both of the two covering situations yield the same result that the V700 absorbing gas covers the whole accretion disk and does not cover BELR (${\rm CF_{V700}^{AD}}=1$ and ${\rm CF_{V700}^{BELR}}=0$), agreeing with what we had guessed from directly comparing H$\alpha$ and H$\beta$ absorption troughs in the SDSS spectrum.
{ This is not surprising because the size of BELR can be several orders of magnitude larger than the accretion disk.}
The size of BELR of LBQS 1206+1052 can be estimated to be $6\times10^{17}$ cm using the radius-luminosity relation of Kaspi et al. (2005) as:
\begin{equation}
R_{\rm BELR} = (22.3\pm2.1)
\left( \frac{L_{5100}}{10^{44}\ {\rm erg\ s}^{-1}} \right)^{0.69\pm0.05} \ {\rm lt-days} ,
\end{equation}
{ where $L_{5100}=3.0\times10^{46}$ erg s$^{-1}$ is the monochromatic luminosity at rest frame 5100 \AA.
The radius $R_\lambda$ at which the disk temperature equals the photon energy, $kT=hc/\lambda_{\rm rest}$ is given by Blackburne et al. (2011) as:}
\begin{equation}
R_\lambda=9.7\times10^{15}
\left( \frac{\lambda}{\mu{\rm m}} \right)^{4/3}
\left( \frac{M_{\rm BH}}{10^9 M_\odot} \right)^{2/3}
\left( \frac{L}{\eta L_{\rm edd}} \right)^{1/3}  {\rm cm} ,
\end{equation}
{ where $L$ is the bolometric luminosity, $\eta$ is the accretion efficiency (assumed to be 0.1) and $L_{\rm edd}$ is the Eddington luminosity.
For LBQS 1206+1052, the black hole mass can be computed to be $1.9\times10^9$ $M_\odot$ using the FWHM of H$\beta$ BEL (FWHM$_{\rm H\beta}\sim7000$ \kms) and the continuum luminosity following Greene \& Ho (2005) as:}
\begin{equation}
M_{\rm BH}=(4.4\pm0.2)\times10^6
\left( \frac{L_{5100}}{10^{44}\ {\rm erg\ s}^{-1}} \right)^{0.64\pm0.02}
\left( \frac{\rm FWHM_{H\beta}}{10^3\ {\rm km\ s}^{-1}} \right)^2  M_\odot ,
\end{equation}
{ and the bolometric luminosity is $8 L_B\sim2.4\times10^{46}$ erg s$^{-1}$ using the bolometric correction from Marconi et al. (2004).
The the size of optical emission region ($\lambda\sim$5000 \AA) can be estimated to be $6\times10^{15}$ cm.}
Considering the two orders of magnitude difference between the sizes of BELR and accretion disk, the covering condition suggests that the size of the absorbing gas is larger than the accretion disk and still far smaller than the BELR.
The best-fitting ${\rm CF_{V1400}^{BELR}}$ is also 1, indicating that the V1400 component covers the whole accretion disk and the whole BELR.
As a result, we fixed the covering conditions and redid the fitting.
The Gaussian centers and widths of all of the three ions, which are listed in Table 4, agree with each other for the V700 component, and those of the He I* and Mg II are also in consistent for the V1400 component.
This indicates that the velocity profiles of different ions are similar, thus we fixed the profile parameters for a better measurement of ionic column densities.
The Balmer absorption of the V1400 component is not detected in all of the spectra, so we obtained an upper limit by assuming that the profile is identical to those of He I* and Mg II.
The resultant ionic column densities are listed in Table 4, and the best-fit model is shown in Figure 7.
As can be seen from the figure, the model reproduces the observed spectra well.

We noticed that the He I* column densities measured from TSpec 201512 and YFOSC 201601 observations differ significantly, especially for the V700 component.
Further more, the He I* column densities measured from NIR seems to be systematically lower than when measured from optical.
This may be due to the contribution of the dusty torus to the continuum in NIR band.
He I* $\lambda$10830 line is at a wavelength where the emission of hot dust reaches its peak, thus it is natural that a part of continuum comes from hot dust.
The emission region of hot dust has a larger size than the accretion disk and the BELR, thus it is likely that the hot dust emission is not covered by the two components.
Therefore the optical depths of He I* $\lambda$10830 may be underestimated for both the two components.
By assuming that the He I* column densities are the same in TSpec 201512 and YFOSC 201601 observations, we estimated that the residual flux of He I* $\lambda$10830 of the V700 component in TSpec 201512 spectrum would be only $\sim$0.05, and it is impossible to make a precise measurement of He I* column density in this situation.
Thus we did not adopt the He I* column density from the TSpec 201512 observation.
We also abandoned the value from the TSpec 201505 observation as the situation is similar.

We plotted the column densities of Mg II, He I* and H(n=2) ions as functions of observing time in Figure 8 and compared them with the V-band light curve.
For the V700 component, the variations of Mg II and H(n=2) show the same pattern that the column densities were the highest in 2003, and then decreased in 2012, and then recovered in 2014 and 2016.
The pattern of He I* is also consistent if adopting the measurements from optical spectra.
For the V1400 component, the pattern of Mg II is similar, while the measurement uncertainties of He I* are too large.
The variability pattern of these ionic column densities is just opposite with that of continuum.

\subsubsection{Photo-ionization model}

Figure 13 in Liu15 shows how the column densities of Mg II and He I* depend on the ionization parameter $U$.
Mg II always decreases with increasing $U$.
He I* decreases with increasing $U$ when the absorbing gas is too thin to reach a ionization front, and increases with increasing $U$ when the gas is thick enough so that the ionization front is well developed.
The ionic column densities are inversely correlated with the V-band continuum in LBQS 1206+1052 for both of the two components, in accordance with the first situation, indicating that both of the two absorbing gases are optically thin.
We noticed that the column densities of He I* and Mg II are insensitive to H density $n_H$ in a wide range, while H(n=2) is sensitive to the density of the gas.
We first investigate the V700 component with additional information from Balmer absorption lines.
We used the photo-ionization code CLOUDY (version 10.00, Ferland et al. 1998) to simulate the ionization process in the V700 absorbing gas.
We assumed a slab-shaped geometry, unique density, homogeneous chemical composition of solar values, and an SED of ionizing continuum of commonly used MF87 (Mathews \& Ferland 1987).
The undetermined parameters are the ionization parameter $U$, the H density $n_H$ and the stop column densities $N_H$.
For a quick look of the general situation, we calculated a sparse grid of models with log$n_H$ varying from 3 to 12, and with log$U$ varying from -3.5 to 2.0, and with log$N_H$ varying from 20 to 24, and the steps of the three are 0.5 dex.
The variable ionization state model assumes that only $U$ is variable among the observations and that $n_H$ and $N_H$ are invariable.
Thus, we check how the three ionic column densities vary as functions of ionization parameter log$U$.
As can be seen in Figure 9, $N_{\rm He I*}$ has an upper limit for a given $N_H$ value, and therefore we can derive a lower limit of the $N_H$ of $\sim10^{21}$ cm$^{-2}$ using the maximum $N_{\rm He\ I*}$ measured from SDSS 200303 observation.

To find the approximate photo-ionization solutions, we plotted the contours of the column densities of He I*, Mg II and H(n=2) according to the levels of observed values in Figure 10.
The observed values are selected from SDSS 200303, YFOSC 201205, DBSP 201404 and YFOSC 201601 observations.
We abandoned the $N_{\rm Mg\ II}$ values from the two YFOSC observations since the uncertainties are too large.
For $N_{\rm H(n=2)}$ in 2012, the value from the MMT 201203 observation was used instead of that from YFOSC 201205 because it has a smaller uncertainty, and because we did not find variation of $N_{\rm H(n=2)}$ between the two epochs which have a rest-frame time interval of only 50 days.
Figure 10(a) shows that the three sets of lines from three ions roughly intersect at a position with log$n_H\sim9$ for every $N_H$ value, supporting the results of former works that Balmer absorption lines are only detected when the density is high.
For log$N_H=21$, the red valid line (contour of $N_{\rm He I*}$ in SDSS 2003 spectrum) does not occur in the region of $n_H\sim9$.
It can be seen from Figure 9 that the peak value of $N_{\rm He I*}$ depends mainly on $N_H$ and slightly on $n_H$, and we found that the red valid line only occurs in the region with log$n_H\sim9$ when log$N_H>21.1$.
For log$N_H=21.5$, the three sets of lines roughly intersect in regions of log$n_H\sim9$ and log$U\sim-1$.
In addition, the intersection points using ionic column densities from four observations have close $n_H$ values (within 0.3 dex), which is in accordance with the variable ionization state model.
For higher $N_H$ values, the $n_H$ values of intersection points from four observations differ by $>$0.5 dex.
This implies that an upper limit of $N_H$ could be obtained under the assumption that $n_H$ is invariable amongst all of the observations.

We then calculated a denser grid of models around the approximate solutions and searched for the photo-ionization solutions.
We assumed that the model has { 6 free parameters, $n_H$, $N_H$ and four $U$.
We compared the difference between spectra predicted by Cloudy model and the observed spectra, and calculated the $\chi^2$ using the difference, and the weights is computed following the method described in Section 3.1.}
Only Balmer and He I* series were used because they have higher measurement accuracies relative to Mg II for the V700 component, and less influenced by metallicity.
The final solution was computed { by minimizing $\chi^2$.}
The best-fitting solution has log$N_H=21.25$, log$n_H=9.38$ and log$U$ varying from $-1.66$ to $-1.29$.
We cut at { $\Delta\chi^2 = 3$} and obtained $21.1<\rm{log}N_H<21.4$ and $9.3<\rm{log}n_H<9.5$, and the solutions by fixing log$N_H$ at 21.1, 21.25 and 21.4 are listed in Table 4 and labeled in Figure 10(b).

\subsubsection{Analysis of HST/COS spectrum}

The density of the V1400 component cannot be determined from the optical and NIR spectra because we only detected He I* and Mg II absorption lines, both of which are insensitive to density (see Figure 9).
Fortunately, there is an archival FUV spectrum of LBQS 1206+1052 taken with HST/COS in 2010, which shows abundant absorption lines in rest-frame 820--1035 \AA.
Chamberlain \& Arav (2015, hereafter CA15) analyzed the HST/COS spectrum.
CA15 noticed that the deepest place in the trough has a velocity of blueshifted $\sim$1400 \kms, which coincides with the V1400 component.
They also found that the absorption lines exhibit a profile that is skewed towards the red side of the trough.
This profile is similar to that of He I* $\lambda$10830, indicating that both of the V1400 and V700 components contribute to the absorption troughs.
CA15 measured the density of the absorbing gas using the line ratios of N III*/N III and S III*/S III.
The two components are blended in N III absorption trough, and meanwhile only V1400 component are detected in the S III absorption trough, thus the density of the V1400 component can be determined as log $n_{\rm H}=3.0\pm0.2$ using S III following CA15.

CA15 also measured the column densities of H I, N III, O III, O VI, S III and S VI and obtained log $U=-1.82$ and log $N_{\rm H}=20.46$ using Cloudy simulation.
Some of the ionic column densities from CA15 are measured by integrating the apparent optical depth (AOD) over the trough, which is mainly contributed by the V1400 component.
Thus, we deduced that the ionic column densities and the photo-ionization solution from CA15 apply to the V1400 component.
{ However, the results may suffer from the contamination by the V700 component.
Thus we need to examined this and measured the physical condition of the V1400 absorbing gas again.}

We attempted to predict the FUV absorption spectra for the two components, in order to better understand the HST/COS spectrum.
We first predicted the ionic column densities of the two components using Cloudy simulation, in which the \nh, \NH\ and $U$ for the V1400 component are from CA15, and those for the V700 component are from our best-fitting solution of log$N_H=21.25$ and log$n_H=9.38$.
The V-band light-curve shows that the quasar luminosity in 2010 is close to that in 2012, therefore the $U$ value from MMT/YFOSC 2012 observation is adopted.
The ionic column densities from Cloudy are then converted to optical depths using oscillator strengths, { which is obtained from NIST Atomic Spectra Database\footnote{http://www.nist.gov/pml/data/asd.cfm}.}
We then generated absorption spectra using Gaussian optical depth profiles obtained in Section 4.1.1.
The results for the two components are shown in Figure 11 for two spectral ranges, along with the observed HST/COS spectrum.
As can be seen in the figure, { most of the absorption troughs can find corresponding features in the predicted absorption spectra.}

However, there are contradiction between the observed and the predicted absorption spectra in the Ly$\beta$ and O VI region (red box in Figure 11).
The photons in the range of $1027<\lambda<1032$ \AA\ are predicted to be fully absorbed by the O VI $\lambda$1031 of the V700 component, while the observed spectrum shows a strong excess.
The excess seems to have a Gaussian shape, and thus we proposed that it may be O VI emission line.
We plotted the observed spectrum in velocity spaces of C III $\lambda$977 and O VI $\lambda$1031 in Figure 12.
The red side of C III emission line is unabsorbed and we can measure the redder position at half maximum at $v\sim600$ \kms, and analysis of profiles of O III and N III emission lines also yield similar results.
On the other hand, by assuming the excess in O VI absorption trough is O VI $\lambda$1031 emission line, the bluer position at half maximum of the emission line profile is at $v\sim-400$ \kms.
Thus the FWHM of the UV emission lines are $\sim$1000 \kms, far narrower than the BELs in optical (FWHM$\sim$6000 \kms), and still broader than the NELs (FWHM$\sim$500 \kms).
A similar situation was seen in OI 287 (Li et al. 2015), where UV emission lines such as Ly$\alpha$ and C IV are narrower than Balmer BELs and broader than [O III] NELs.
This can be explained by introducing additional intermediate-width emission lines that originate in the inner face of the dusty torus, and by assuming that the UV continuum and UV BELs are suppressed by dust extinction.
We plotted the SED of LBQS 1206+1052 in UV band in Figure 13.
The YFOSC 2012 spectrum was used to generate the SED together with HST/COS spectrum because the light curve shows that the quasar luminosities in the two epochs are close.
We found that the continuum in the two spectra can be well fitted by a reddened quasar composite spectrum from VanDen Berk (2001) by an SMC extinction curve with $E_{B-V}=0.07$.
This implies that the continuum in the HST/COS spectrum is dust extinguished by about one order of magnitude.
Therefore it is likely that the emission lines in the HST/COS spectrum of LBQS 1206+1052 are from intermediate-width emission line regions, just as in OI 287.

{ CA15 showed their Cloudy simulation results in their Figure 3.
They determined the $U$ and $N_H$ using the cross point of O VI contour and other ions in $U$--$N_H$ space.
We argued that the apparent residual flux in O VI trough is likely to be an emission line, thus the measured O VI column density may not be real.
Then how to determine the physical properties of V1400 absorbing gas?
We noticed that S VI contour also crosses with other ions.}
As can be seen in Figure 11 (blue box), the blue side of S VI $\lambda$944 for the V1400 component is not affected by the V700 component.
Thus we measured the S VI column density by integrating the AOD in the blue side and multiplying by two.
{ Now we can remeasure $U$ and $N_H$ using a Cloudy simulation.
The settings of the Cloudy simulation are similar with those for the V700 component, except that $n_H$ is fixed at $10^3$ cm$^{-3}$.}
The ionic column densities measured by us are listed in Table 5, along with the values from CA15.
{ The absorption troughs of O III, C III and N III are deep, and the measured column densities may be underestimated due to unknown flux which is not covered by the absorbing gas, such as scatter light and emission lines.}
Thus we did not adopt the column densities of these ions when searching for photo-ionization solution.
CA15 measured the H I column density from Lyman-limit measurements, which is contaminated by V700 component, and we did not adopt the value either.
In addition, we detected weak Ar IV $\lambda\lambda$844, 850 and C II $\lambda$903 absorption lines for the V1400 component (purple box in Figure 11), and the column densities of Ar IV and C II are also measured.
The Cloudy simulation results are plotted in Figure 14(a) and the best-fitting solution is labeled using black star.
The ionic column densities predicted by the solution are listed in Table 5.
As shown in the table, the observed and predicted column densities of S VI, S III, Ar IV and C II agree within 0.3 dex.

Relative to CA15's solution (triangle in Figure 14(a)), our solution has a $U$ 0.22 dex higher and an $N_H$ 0.47 dex higher.
As the contours of most ions are in lower-left to upper-right direction in $U$--$N_H$ space, ions with higher ionization energies, such as O VI, S VI and Ar IV, are important to determine the solution because their contours cross with the contours of low ionization ions.
In C15's simulation, the cross point of S VI is different than the cross point of O VI.
If adopting the cross point of S VI in CA15's simulation, the solution (square in Figure 14(a)) is much closer to ours, and the difference with $<$0.1 dex is due to applying a different SED.
The cross point of Ar IV is close to that of S VI, making the solution of using S VI more reliable.

The HST/COS spectrum shows an additional absorption line system with a blueshift velocity of $\sim$200 \kmsb (hence is referred to as the V200 component hereafter), which is seen in Lyman series, C III $\lambda$977 and N III $\lambda$989, and the corresponding absorption lines are labeled using pink dashed lines in Figure 11.

\subsubsection{Consistency between variations in continuum and in U}

The variable ionization state model assumes that all of the other parameters are invariable and thus the variation of $U$, which is presented as $U=\frac{Q(H^0)}{4\pi r^2 c n_H}$, is proportional to the variation of continuum.
We showed that the observed ionic column densities are negatively correlated with the optical continuum flux.
The photoionization simulation indicates that the ionic column densities of He I*, Mg II and H(n=2) are all negatively correlated with $U$.
These are consistent.
The photoionization simulation presented the change in $U$ as 0.29--0.43 dex between the SDSS 2003 and YFOSC 2012 observations.
The corresponding variability amplitude of the optical continuum is 0.26 mag ($\sim$0.1 dex) in the V-band.
Considering that quasars are often more variable in UV and X-ray than in optical, the variation in the flux of the ionization continuum would be larger.
By comparing the GALEX photometry in 2004 with the reddened quasar composite which can join the HST/COS 2010 and YFOSC 2012 spectra (Figure 14), we estimated the variability amplitude between 2004 and 2010 to be 0.1 and 0.15 dex in the FUV and NUV band.
The FUV band is strongly influenced by the Ly$\alpha$ emission line whose EW may be different in LBQS 1206+1052 and in quasar composite, thus the variability amplitude of NUV was adopted.
This finding supports the idea that the variability in the ionization continuum is responsible for the variation in $U$ in LBQS 1206+1052.

Using the $U$ and $n_H$ values from the photo-ionization simulations, we estimated the distance $r$ of the absorbing gases to the central source.
We obtained $Q(H^0)$ to be $7.4\times10^{56}$ s$^{-1}$ by integrating the MF87 SED which is normalized using the observed $L_{3000} = 4.7\times10^{45}$ erg s$^{-1}$ in the YFOSC 2012 spectrum.
The far-UV continuum luminosity from HST/COS spectrum was not used it is probably under heavy dust extinction.
For the V700 absorbing gas, the distance was estimated to be $1.3\pm0.4$ pc using the $U$ value in 2012, which is corresponding to the $Q(H^0)$ in 2012, and for the V1400 absorbing gas the inferred distance is $\sim$2.8 kpc.

\subsection{Variable Covering Factor Model}

We found observational evidence against the variable covering factor model.
In this model the optical depths of all of the absorption lines are invariable, and thus the change in EW is proportional to the change in the covering factor.
The EW of He I* $\lambda$3889 for the V700 component in the YFOSC 201601 observation is $0.54^{+0.16}_{-0.20}$ (90\% confidence) times the EW value in the SDSS 200303 observation.
Assuming that the V700 absorbing gas covers the whole continuum in the SDSS 200303 observation, we can estimate the upper limit of the covering factor in YFSOC 201601 to be 0.7.
On the other hand, another covering factor of He I* can be obtained from the He I* $\lambda$10830 absorption trough in the quasi simultaneous 201512 TSpec observation.
It can be seen from Figure 6 that the absorbing gas covers at least 80--90\% of the continuum.
This fraction may be higher considering the underneath hot dust emission.
Therefore the covering factors measured from the two observations with only 8 days interval in rest frame are not consistent if assuming the ionic column density is invariable.
We have estimated the size of the accretion disk to be $\sim10^{15}$ cm.
This requires a crossing velocity of $\sim14000$ \kms, which is too large for an absorbing gas with a radial velocity of only 700 \kms.
Moreover, if the absorbing gas moved across in such a short time, it is difficult to explain how the absorption troughs remained over ten years.

The variation in absorption lines for the V1400 component is also strange under the assumption of this model.
The covering factor of Mg II $\lambda$2796 of the V1400 component to the continuum increased about twice between YFOSC 201205 and DBSP 201404, while the covering factor of He I* $\lambda$10830 changed little between 2013 and 2015.
One can only explain this by introducing more complicated models, such as adopting different covering factors for He I* and Mg II.
However, the velocity profiles of He I* and Mg II are nearly identical, indicating that the two ions are located in the same position, and hence it is likely that the covering factors are also the same.

The model is also difficult to understand kinetically.
If the weakening of absorption lines from 2003 to 2012 is due to movement of an absorbing gas cloud, the recovery in 2014 is unusual because the cloud would not move back.
If the recovery is due to the entering of a new cloud, it is difficult to explain why the velocity profiles of the two clouds show no difference.
As we have reproduced all of the data using the variable ionization state model, we excluded this variable covering factor model.
We conclude that the absorption line variability in LBQS 1206+1052 is due to a change in ionization state.

\section{Discussion}

\subsection{Robustness on the parameters of the two components}

We first discuss the parameters of the V700 component.
The lower limit of $N_H$ (log$N_H>21.1$) was from the observed $N_{\rm He\ I*}$, which is rather robust.
The upper limit of log$N_H<21.4$ was derived by assuming that $n_H$ is invariable amongst the observations.
In the process we only considered the measurement uncertainty while the systematic errors in the photo-ionization simulation are not accounted for.
Considering the systematic errors, the possible ranges of the parameters would be larger.
The influence of uncertain metallicity is little because the BAL parameters are determined using He I* and H(n=2), both of which are insensitive to metallicity.
In addition the $N_{\rm Mg\ II}$ from best-fit solution and from the observations are consistent by 0.3 dex, suggesting that the chemical composition of the V700 absorbing gas is similar to that of the solar one.
On the other hand, the SED of the ionization continuum indeed affects the measurement of the parameters.
We tested another SED, which is described in detail in Ji et al. (2015), and has a lower flux in extreme UV and X-ray than the MF87 SED if normalized at 3000 \AA.
Using $N_{\rm He\ I*}$ the lower limit of log$N_H$ is 21.3, slightly larger than using MF87 SED.
The upper limit is also larger with log$N_H<22.5$, and the $N_H$ range is also wider.
The $n_H$ is in the range of $9.4<\textrm{log}n_H<9.9$, and the inferred distance is in the range of 0.2 to 0.3 pc using a lower $Q(H^0)$ of $3.3\times10^{56}$ s$^{-1}$.
There is no available X-ray data for LBQS 1206+1052 and the UV band suffers from dust reddening, and thus we have no observational limitation to the SED of ionization continuum.
Thus the uncertainty of parameters due to uncertain SED should also be considered.
Combining the results by using different SEDs, we concluded that the $n_H$ is in the range of $10^9$--$10^{10}$ cm$^{-3}$, and the distance of the absorbing gas to the central source is in the range of 0.2 to 2 pc.

We then discuss the parameters of the V1400 component.
The $n_H$ is converted from $n_e$, which is measured using the ratio between the excited and ground states of [S III].
[S III] absorption lines are optically thin and hence do not suffer from uncertainty in unknown covering conditions, yet still have a high enough S/N ratio to ensure an accurate measurement.
Thus the measurement of $n_H$ is reliable.
The $N_H$ and $U$ values are from Cloudy simulations.
As CA15 showed, the photoionization solution is independent of the metallicity of the gas, and the choice of SED
can influence the measurement of $U$ by up to 0.3 dex.

{ Since the variability is photo-ionization driven, the density can also be limited using the recombination time scale $\tau=(n_e \alpha_r)^{-1}$, where $\alpha_r$ is the recombination rate, which is related with the electron temperature.
For a typical ionized gas, $T\sim10^4$ K, and thus the recombination rates are $4.2\times10^{-13}$, $4.6\times10^{-13}$ and $1.2\times10^{-12}$ cm$^3$ s$^{-1}$ for H I, He I and Mg II, respectively, which were obtained from Verner \& Ferland (1996).
For the V700 absorbing gas with $n_H\sim10^{9.4}$, the recombination time scale of the three ions are all less than $10^4$ seconds assuming $n_e=1.2n_H$ for a highly ionized gas.
Therefore the variation of ion column densities can trace the continuum variability immediately for the V700 component.
For the 1400 component, we can similarly estimated a recombination time scale to be $\sim$20 years using $n_e=10^{3.0}$ cm$^{-3}$, too long to match the observation.
And by assuming the recombination time scale is $<$1 year, the inferred $n_e>10^{4.4}$ cm$^{-3}$, higher than that measured from S III.
We noticed that for V1400 component, only the response of Mg II to continuum can be seen, and we did not see the variability of He I* and other ions.
Considering the ionization energy of Mg II (7.6 eV) is much lower than S III (23.3 eV) and other ions, a possible explanation for the contradiction is that Mg II comes from a region with higher density, and thus has a shorter recombination time scale.
This requires a two-phase medium for the V1400 absorbing gas, and the parameters we obtained may apply to the lower density phase.}

The distance of the V1400 absorbing gas is three orders of magnitude larger than the V700 absorbing gas.
Though the measurements of distances have uncertainty, we can conclude that the V1400 absorbing gas is located behind V700 the absorbing gas relative to the quasar.
Thus we needed to consider that the ionization continuum of the V1400 absorbing gas is filtered by the V700 absorbing gas.
A Cloudy simulation shows that the V700 absorbing gas causes a total decrease in $Q(H^0)$ of 40\% and that the output SED is also different than the input SED, as nearly all of the photons with $54.6<E<100$ eV are absorbed (see Figure 14(b)).
We referred to this effect as ``shading effect'' hereafter.
We made another photo-ionization simulation of the V1400 absorbing gas using the transmitted spectrum of the V700 absorbing gas as the input SED.
The shading effect mainly causes a decrease of column densities of high-ionization ions, such as S VI, O VI and Ar IV, and the low-ionization ions are less affected.
The solution is at log$U=-1.13$ and log$N_H=21.09$ (red star in Figure 15), which moves towards the up-right direction in the $U$--$N_H$ space relative to the solution without the shading effect.
Considering the shading effect, the $U$ value from the Cloudy simulation is higher and the ionization photons are reduced.
Using the effective $Q(H^0)$ of $4.4\times10^{56}$ s$^{-1}$, the best estimation of the distance of the V1400 absorbing gas is about 1.2 kpc.

\subsection{The absorbing gases and outflows in LBQS 1206+1052}

We showed that there are three absorption line systems in the spectra of LBQS 1206+1052 with blueshifted velocities of $1420$, $700$ and $200$ \kmsb relative to the quasar rest frame, respectively.
The V700 absorbing gas has an $n_H$ in the range of $10^9$ to $10^{10}$ cm$^{-3}$.
It is located at a distance of $\sim$1 pc from the central source, just outside of the BELR.
The virialized velocity in this distance is $v\sim\sqrt{\frac{GM_{\rm BH}}{r}}=2000$ \kms, and thus the gas may be restricted in this region.

The V1400 absorbing gas is at a distance of $\sim$1.2 kpc and with an $n_H$ of $10^3$ cm$^{-3}$.
The blueshifted velocity of $\sim1400$ \kmsb indicates it is an outflow, as it is in line of sight of the quasar.
We found a strong blue wing in the profiles of NELs of [O III] $\lambda\lambda$4959, 5007 and [Ne III] $\lambda$3869, which is extended to a velocity of $-2500$ \kms.
We measured the luminosity of [O III] $\lambda$5007 and [Ne III] $\lambda$3869 in a velocity range from $-2000$ to $-1000$ \kmsb to be $6\times10^8$ $L_\odot$ and $6\times10^7$ $L_\odot$, respectively.
The distance and density of the V1400 absorbing gas are both similar with NELR, where the blue wing of NELs may originate in, and the blueshifted velocities of the V1400 absorbing gas and the blue wing are also consistent.
These similarities suggest a link between the blue wing of NELs and the V1400 absorbing gas.
Thus we calculated the outflow rate in the kpc scale by assuming that the outflow seen in the blue wing is formed by a layer of gas clouds with similar distances and properties with the V1400 absorbing gas.
For this purpose we first simulated the emission line spectrum of the layer by assuming a spherical shell of gas with radius of 1 kpc, $n_H$ of $10^3$ cm$^{-3}$ and radial H column density of $10^{21.09}$ cm$^{-2}$.
We used MF87 SED instead of the transmitted SED of the V700 absorbing gas.
The results show that the layer of outflowing gas can produce enough of the [O III] and [Ne III] emission lines measured above if the global covering factor is $>10^{-3}$, and the mass outflow rate is estimated to be $0.3 M_\odot$ yr$^{-1}$, and the inferred kinetic outflow rate is $1.5\times10^{41}$ erg s$^{-1}$.
The total mass outflow rate in this distance may be several times higher considering those gases with lower or higher velocities, and the possible gases in the opposite direction, which are probably obscured by dusty torus.
The estimation assumes that all of the gases in this region have the same density with the V1400 absorbing gas, and considering the typical density in NELR of $10^2$--$10^4$ cm$^{-2}$, the mass outflow rate may have uncertainty of one dex.

The V200 absorbing component shows too few lines and we cannot obtain a good estimation on the properties of the absorbing gas.
We attempted to set a constraint on its properties using the non-detection of several lines.
The non-detection of N III $\lambda$991 can set an upper limit of the $n_H$ as $10^2$ cm$^{-3}$.
The non-detection of the high-ionization line S VI $\lambda$944 suggests either a lower $U$ in the V200 absorbing gas than in the V1400 absorbing gas, or a further decrease in hard photons due to the shading effect of the V1400 absorbing gas, both of which indicate a larger distance of the V200 absorbing gas relative to the V1400 absorbing gas.
The lower blueshift velocity also implies that V200 absorbing gas is in an outer region.

We found extended emission line regions (EELRs) around LBQS 1206+1052 from the long slit optical spectra.
The EELRs were found in MMT 201203, YFOSC 201205, DBSP 201404 and YFOSC 201601 spectra with different position angles of 21, 0, 24 and 90 degrees, respectively.
[O III] $\lambda\lambda$4959, 5007, H$\alpha$, H$\beta$, [Ne III] $\lambda$3869 and [O II] $\lambda$3727 emission lines were seen for the EELRs.
The highest flux of the extended emission line feature was recorded in the YFOSC 2012 spectrum with a long slit in the N--S direction.
We plotted the two-dimensional spectral image in Figure 16.
The brightest position is at 2.8\arcsec\ (15 kpc in projection) north to the quasar and the EELR extends for a maximum distance of $\sim$5\arcsec.
After subtracting the contribution of traditional NLR, the measured extended [O III] luminosity is $1.1\times10^{43}$ erg s$^{-1}$, and the total luminosity is higher considering this luminosity is only extracted from a slit with 1.8 arcsec width.
This luminosity is very high among the EELRs, e.g. the highest extended [O III] luminosity recorded in Fu \& Stockton (2009) is $1.0\times10^{43}$ erg s$^{-1}$.
Assuming the density of the EELR is similar to those measured by Fu et al. (2009), the mass of the gas can be estimated to be $3\times10^6\ M_\odot$ using H$\beta$ luminosity.
The extended [O III] feature shows radial velocities difference relative to quasar rest-frame in both sides, and we measured a redshift velocity of $\sim100$ \kmsb on the North side.
We have seen outflows in the kpc scale with enough mass and a high enough velocity to form a structure with a size of 15 kpc.
If the EELR was formed by a similar outflow in the past, the dynamic time scale can be estimated to be $\sim1.5\times10^8$ yrs using the size and an expanding velocity of $\sim$100 \kms, which is close to the lifetime of AGNs (e.g., Marconi et al. 2004).
Therefore we presented an overall scenario of outflows in LBQS 1206+1052 that the quasar started to drive outflows when it was born $\sim10^8$ years ago, and continuously generated strong outflows until at least $\sim10^6$ years ago, which is the dynamic time scale of the V1400 outflow.
A detailed analysis of the EELR is beyond the scope of this work.
Future integral field spectrographic observations are needed to better understand the large scale outflow in LBQS 1206+1052.

\subsection{Future follow-up observations of He I* and Balmer BAL quasars}

Changes in ionization state and movement of absorbing gas are the best two origins of absorbing line variability.
To date, no one can unambiguously say which is more important amongst BAL quasars.
He I* and Balmer multiplets are powerful in directly determining covering conditions and measuring ionic column densities.
Therefore follow-up observations of He I* BAL quasars and Balmer BAL quasars can distinguish the two possible origins in these quasars, and examples are presented in Shi16 and this work.
He I* absorption lines exist in nearly all LoBALs quasars, thus follow-up observations of a sample of He I* BAL quasars can tell us the fraction of those that show variable ionic column densities and those that show variable covering factors.
In this way we can make clear the origin of absorption line variability in LoBAL quasars.

{ Balmer BAL quasars are rare but important because it trace high density media.
We need to measure the properties of absorbing gases in Balmer BAL quasars to better understand this rare type of BAL and high density media around quasar nuclei.
Typically column densities of three ions are needed to obtain the basic parameters, such as $n_H$, $N_H$ and $U$.
For example, objects studied by Zhang et al. (2014) and Shi et al. (2016a, b) are iron LoBAL, and they used Balmer, He I* and excited states of Fe II to calculate all the three parameters.
We show in this work that by taking follow-up observations, we can limit the parameters using only Balmer and He I*.
Considering that all known Balmer BAL show corresponding He I*, this means that the density and position of absorption media can be determined for most of the Balmer BAL quasars.
Thus we propose that follow-up observations of Balmer BAL quasars are essential to better understand this rare population, especially those are not iron LoBAL.}

\section{Summary}

We present the analysis of absorption variability in mini-BAL quasar LBQS 1206+1052.
We first recovered the absorption-free spectra using a pair-matching method.
After normalization, we found an anti-correlation between the variation in EW of absorption lines and the V-band light curve, suggesting that the variability is photoionization driven.
We then attempted to reproduce the data using two models; one assuming that only the ionization state is variable and that all the other physical properties are invariable, and the other assuming that only the covering factor is variable.
In the variable ionization state model, the two components are both optically thin and thus the column densities of He I*, Mg II and H(n=2) all decrease with increasing ionization parameter, demonstrating the anti-correlation between the EW of absorption lines and the continuum flux.
The ionic column densities from the best-fit photoionization solution are also in agreement with the measured values, and the variation of $U$ in the model is also in agreement with the observed variation in continuum.
In this way we present that the variable ionization state model can successfully reproduce the data.
On the other hand, the variable covering factor model can not reproduce the data because of the inconsistency between covering factors of He I* from quasi-simultaneous optical and NIR spectra, the unsynchronized variation between He I* $\lambda$10830 and Mg II during 2012 and 2015, and the behavior of first weakening and then strengthening which is difficult to understand kinetically.
Therefore we conclude that the absorption line variability in LBQS 1206+1052 is photoionization driven.

We also obtained the physical properties of the two absorbing gases.
The V700 absorbing gas has a size similar to the accretion disk, and is at a distance of $\sim$1 pc to the central source, and has a density in the range of $10^9$ to $10^{10}$ cm$^{-3}$.
The velocity and position suggests that it is a restricted structure just outside the BELR.
The V1400 absorbing gas represents an outflow with a distance of $\sim$1 kpc, and with a density of $10^3$ cm$^{-3}$.
We related the outflow with the blue wing of the NELs and estimated a mass outflow rate of $>0.3\ M_\odot$ yr$^{-1}$ and a kinetic outflow rate of $>1.5\times10^{41}$ erg s$^{-1}$ at this distance.
We suggest that the large scale outflow started at $\sim10^8$ years ago as we found an EELR, which is worth follow-up investigations.

\clearpage

\begin{table}[!t]\footnotesize
\caption{Summary of spectral observations of LBQS 1206+1052}
\begin{threeparttable}
\begin{tabular}{cccccccccc}
\hline
\hline
Instrument & obs-date & exposure & rest-frame wavelength  & SNR\tnote{a} &\multicolumn{5}{c}{resolutions\tnote{b}}\\
           &          & (s)      & (\AA)                  &      &2800 &3889 &4862 &6563 &10830\\
\hline
SDSS            &2003/03/24 &3165  &2720--6608                      &46 &1550 &2050 &1900 &2400 &-\\
HST/COS         &2010/05/08 &4840  &819--923 \& 931--1034\tnote{c}  &-  &\multicolumn{5}{c}{16000--21000}\\
MMT/Red Channel &2012/03/01 &500   &6284--6823                      &-  &-    &-    &-    &3500 &-\\
Lijiang/YFOSC   &2012/05/07 &4800  &2410--5540                      &46 &360  &500  &640  &-    &-\\
P200/TripleSpec &2013/02/23 &720   &6970--17650                     &34 &-    &-    &-    &-    &2200\\
P200/DBSP       &2014/04/24 &600   &2262--4191 \& 5616-7673\tnote{d}&55 &1000 &1300 &-    &2700 &-\\
P200/TripleSpec &2015/05/26 &1200  &6970--17650                     &37 &-    &-    &-    &-    &2200\\
P200/TripleSpec &2015/12/28 &540   &6970--17650                     &12 &-    &-    &-    &-    &2200\\
Lijiang/YFOSC   &2016/01/08 &2590  &2410--5540                      &23 &300  &420  &520  &-    &-\\
\hline
\hline
\end{tabular}
\begin{tablenotes}
    \item [a] The signal to noise per Angstrom, which is measured around 3800 \AA\ for SDSS, DBSP and YFOSC, and around 12000 \AA\ for TripleSpec.
    \item [b] The spectral resolution R = $\lambda / {\rm FWHM}$. For the 8 optical and NIR observations, we listed the resolutions at rest-frame wavelengths of 2800, 3889, 4862, 6563 and 10830 \AA, corresponding to Mg II, He I* $\lambda$3889, H$\beta$, H$\alpha$ and He I* $\lambda$10830 regions.
    \item [c] The wavelength ranges of the two CCD chips.
    \item [d] The wavelength ranges of the blue and red sides of DoubleSpec.
\end{tablenotes}
\end{threeparttable}
\label{tab1}
\end{table}

\begin{table}[!t]\footnotesize
\caption{The EW ratios of optical absorption lines in YFOSC and DBSP spectra relative to in SDSS spectrum and uncertainties with 90\% confidence.}
\begin{threeparttable}
\begin{tabular}{c|cccc}
\hline
\hline
 &\multicolumn{4}{c}{EW/EW$_{\rm SDSS}$}\\
 &Mg II doublet           &He I $\lambda$3889 &H$\alpha$   &H$\beta$ \\
velocity range   &$-3000<v<0$ (for 2803)  &$-1000<v<0$        &$-1400<v<0$ &$-1400<v<0$ \\
\hline
MMT 201203      &-                      &                       &$0.19^{+0.16}_{-0.16}$ &\\
YFOSC 201205    &$0.24^{+0.07}_{-0.12}$ &$0.16^{+0.21}_{-0.20}$ &-&$0.47^{+0.18}_{-0.16}$ \\
DBSP 201404     &$0.89^{+0.11}_{-0.15}$ &$0.62^{+0.15}_{-0.14}$ &$0.69^{+0.14}_{-0.24}$ &\\
YFOSC 201601    &$0.83^{+0.10}_{-0.13}$ &$0.54^{+0.16}_{-0.20}$ &-&$0.61^{+0.38}_{-0.41}$ \\
\hline
\hline
\end{tabular}
\end{threeparttable}
\label{tab2}
\end{table}

\begin{table}[!t]\footnotesize
\caption{The best values and 90\% confidence intervals of the absorption line parameters.}
\begin{threeparttable}
\begin{tabular}{c|ccc|ccc}
\hline
\hline
&\multicolumn{3}{|c|}{V700 component} &\multicolumn{3}{|c}{V1400 component}\\
 &Mg II  &He I*  &H(n=2)   &Mg II  &He I*  &H(n=2) \\
\hline
\multicolumn{7}{c}{line profiles}\\
\hline
line center         &$-671^{+66}_{-76}$ &$-720^{+41}_{-48}$ &$-705^{+46}_{-49}$ &$-1412^{+11}_{-12}$ &$-1417^{+17}_{-18}$ &-\\
line width          &$285^{+25}_{-33}$ &$281^{+26}_{-46}$ &$276^{+31}_{-59}$ &$99^{+6}_{-8}$ &$105^{+21}_{-12}$ &-\\
line center (fixed) &\multicolumn{3}{|c|}{$-698^{+32}_{-33}$}  &\multicolumn{3}{|c}{$-1413^{+13}_{-13}$}\\
line width (fixed)  &\multicolumn{3}{|c|}{$270^{+26}_{-28}$}  &\multicolumn{3}{|c}{$107^{+13}_{-11}$}\\
\hline
\multicolumn{7}{c}{column densities ($10^{13}$ cm$^{-2}$)}\\
\hline
SDSS 200303  &$14.6^{+6.5}_{-3.0}$ &$33.3^{+17.3}_{-5.3}$ &$6.7^{+1.8}_{-2.4}$ &$5.3^{+0.8}_{-0.8}$ &$9.6^{+3.3}_{-3.9}$ &$<0.2$\\
MMT 201203   &-&-&$1.2^{+1.1}_{-0.7}$    &-&-&$<0.05$\\
YFOSC 201205 &$2.5^{+2.1}_{-1.0}$  &$6.8^{+8.8}_{-3.4}$    &$2.2^{+2.9}_{-1.6}$ &$1.5^{+0.5}_{-0.4}$ &$6.0^{+8.8}_{-5.7}$ &$<1.1$\\
TSpec 201302 &-&$<3.7$&-                 &-&$2.6^{+0.8}_{-0.5}$&-\\
DBSP 201404  &$12.2^{+9.7}_{-3.7}$ &$19.6^{+13.1}_{-4.5}$  &$4.3^{+2.2}_{-2.4}$ &$3.6^{+0.9}_{-1.0}$ &$8.0^{+3.3}_{-3.5}$ &$<0.2$\\
TSpec 201505 &-&$5.4^{+4.0}_{-4.3}$&-    &-&$3.8^{+1.0}_{-0.7}$&-\\
TSpec 201512 &-&$3.7^{+3.1}_{-2.1}$&-    &-&$4.0^{+1.4}_{-1.0}$&-\\
YFOSC 201601 &$13.5^{+7.0}_{-4.2}$ &$16.3^{+11.7}_{-5.5}$   &$3.3^{+3.8}_{-2.6}$ &$1.7^{+1.2}_{-0.9}$ &$8.9^{+12.5}_{-7.1}$ &$<1.2$\\
\hline
\hline
\end{tabular}
\end{threeparttable}
\label{tab3}
\end{table}

\begin{table}[!t]\footnotesize
\caption{The parameters of the absorbing gases obtained from Cloudy simulations}
\begin{threeparttable}
\begin{tabular}{c|c|ccc|cc}
\hline
\hline
\multicolumn{2}{c|}{} &\multicolumn{3}{c|}{V700 absorbing gas} &\multicolumn{2}{c}{V1400 absorbing gas}\\
\multicolumn{2}{c|}{} &\multicolumn{3}{c|}{}       &without shading &with shading\\
\hline
\multicolumn{2}{c|}{log$N_H$ (cm$^{-2}$)} &21.1 &21.25 &21.4 &20.93 &21.09\\
\multicolumn{2}{c|}{log$n_H$ (cm$^{-3}$)} &9.32 &9.38  &9.52 &\multicolumn{2}{c}{3.0}\\
\hline
\multirow{5}{0.6cm}{log$U$\tnote{a}}&SDSS 200303 &$-1.9$ &$-1.66$ &$-1.41$ &\\
 &HST/COS 2010    &        &         &        &$-1.60$ &$-1.13$\\
 &MMT\&YFOSC 2012 &$-1.47$ &$-1.29$  &$-1.12$ &\\
 &DBSP 201404     &$-1.77$ &$-1.49$  &$-1.25$ &\\
 &YFOSC 201601    &$-1.65$ &$-1.4$   &$-1.2$  &\\
\hline
\multicolumn{2}{c|}{$\Delta$log$U$(2003 to 2012)} &0.43 &0.35 &0.29 &&\\
\hline
\multicolumn{2}{c|}{distance (pc)}  &1.7&1.3&0.9&2800&1200\\
\hline
\hline
\end{tabular}
\begin{tablenotes}
    \item [a]
\end{tablenotes}
\end{threeparttable}
\label{tab4}
\end{table}

\begin{table}[!t]\footnotesize
\caption{Ionic column densities of V1400 component}
\begin{threeparttable}
\begin{tabular}{ccccc}
\hline
\hline
ion    &\multicolumn{2}{c}{log$N_{ion}$ (cm$^{-2}$)\tnote{a}} &\multicolumn{2}{c}{log$N_{model}$ (cm$^{-2}$)\tnote{b}}\\
       &this work     &CA15    &without shading  &with shading\\
\hline
H I   &-              &16.90--17.06   &17.27  &17.04\\
C II  &$14.0\pm0.3$   &-              &14.35  &14.10\\
C III &-              &$>$14.95       &16.62  &16.80\\
N III &-              &$15.68\pm0.1$  &16.10  &16.22\\
O III &-              &$16.26\pm0.15$ &17.22  &17.76\\
O VI  &-              &$15.55\pm0.1$  &15.94  &14.17\\
S III &$15.1\pm0.2$   &$15.09\pm0.2$  &14.84  &14.79\\
S VI  &$15.22\pm0.1$  &$15.27\pm0.1$  &15.24  &15.18\\
Ar IV &$15.19\pm0.3$  &-              &15.02  &15.40\\
He I* &$13.65\pm0.3$\tnote{c} &-      &13.61  &14.27\\
Mg II &13.23--13.80\tnote{d}  &-      &13.52  &13.84\\
\hline
\hline
\end{tabular}
\begin{tablenotes}
    \item [a] The sum of ground and excited level for each ion.
    \item [b] The column densities predicted by our best-fitting CLOUDY model.
    The results are from two simulations, one not considering shading effect (using MF87 SED) and the other considering shading effect (using transmitted SED of V700 absorbing gas).
    \item [c] The value is from three TSpec observations.
    \item [d] The upper and lower limits are given using the measurements from SDSS 2003 and YFOSC 2012 spectra.
\end{tablenotes}
\end{threeparttable}
\label{tab6}
\end{table}

\clearpage

\begin{figure}
\centering{
 \includegraphics[scale=0.98]{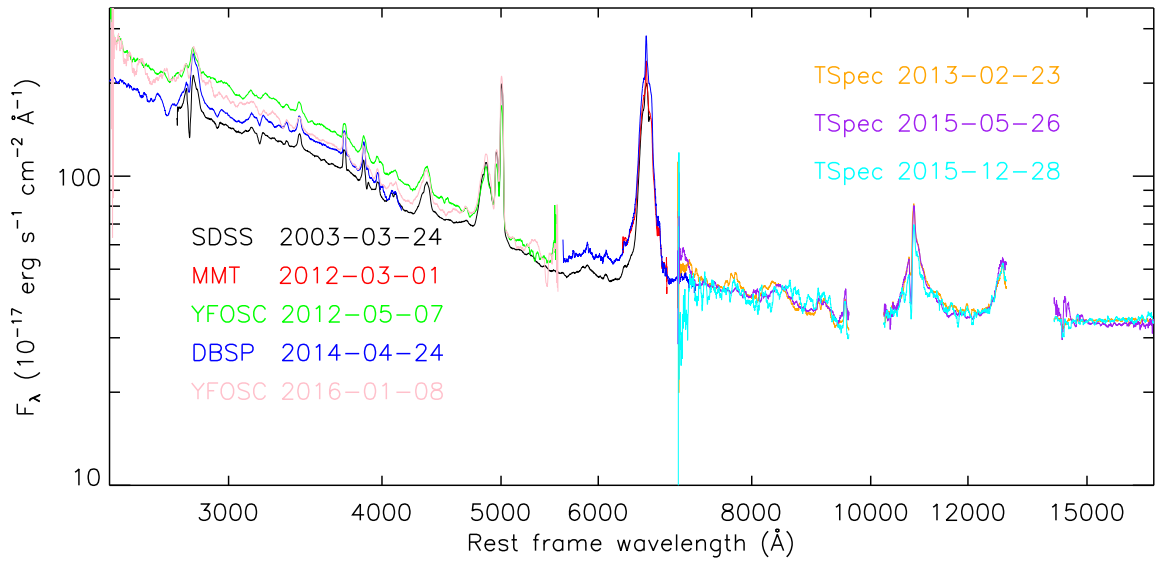}
 \caption{:
  The snapshots of the five optical spectra and three NIR spectra of LBQS 1206+1052.
  The observed wavelength are converted rest-frame vacuum wavelength using $z=0.3953$.
  The instruments and the dates of the observations are labeled.
  }}
\label{allspec}
\end{figure}

\begin{figure}
\centering{
 \includegraphics[scale=0.98]{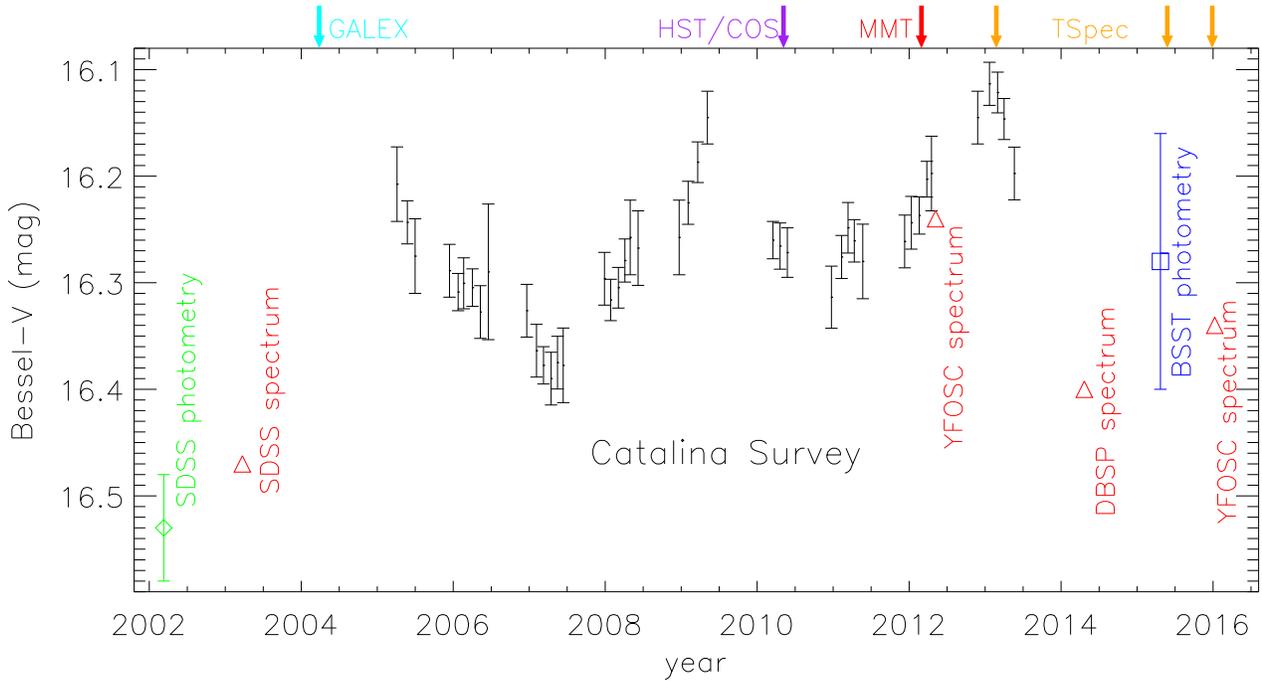}
 \caption{:
  The Bessel-V band lightcurve of LBQS 1206+1052.
  The black data points show the binned light-curve from Catalina Sky Survey, and the green data point shows the magnitude which is converted from SDSS photometry, and the blue data point shows the BSST photometry.
  The 1$\sigma$ error bar are also plotted for the above three sets of photometric data.
  The red triangles represent the synthetic V-band magnitudes from SDSS, YFOSC and DBSP spectra.
  And the observation time of GALEX photometry, HST/COS spectrum, MMT spectrum and three TSpec spectra are labeled on the top.
  }}
\label{lightcurve}
\end{figure}

\begin{figure}
\centering{
 \includegraphics[scale=0.98]{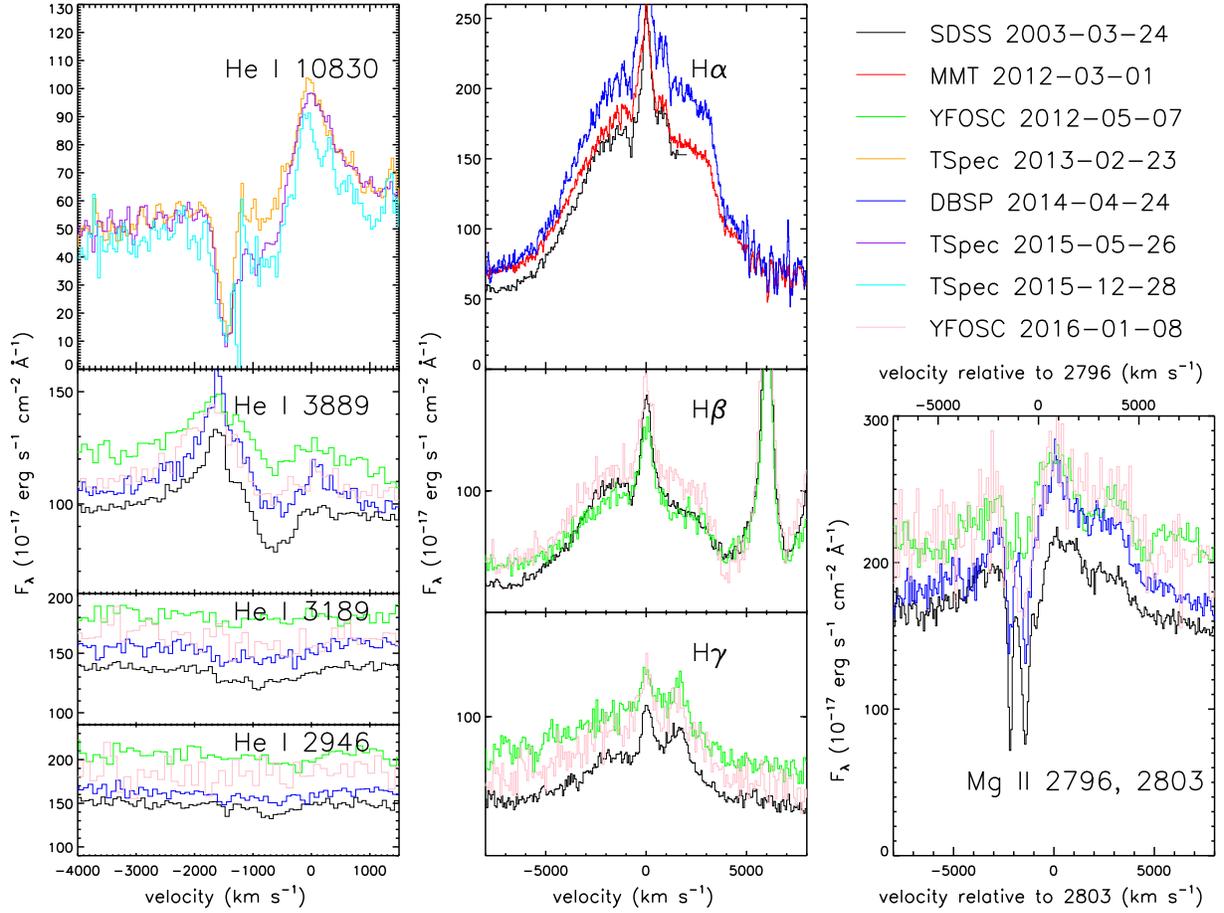}
 \caption{:
  The spectra in absorption windows of He I* $\lambda$10830, $\lambda$3889, $\lambda$3189, $\lambda$2946, H$\alpha$, H$\beta$, H$\gamma$ and Mg II $\lambda$$\lambda$2796,2803.
  The corresponding color for each spectrum is same with Figure 1 and is labeled in the upper-right.
  }}
\label{absorb}
\end{figure}

\begin{figure}
\centering{
 \includegraphics[scale=0.98]{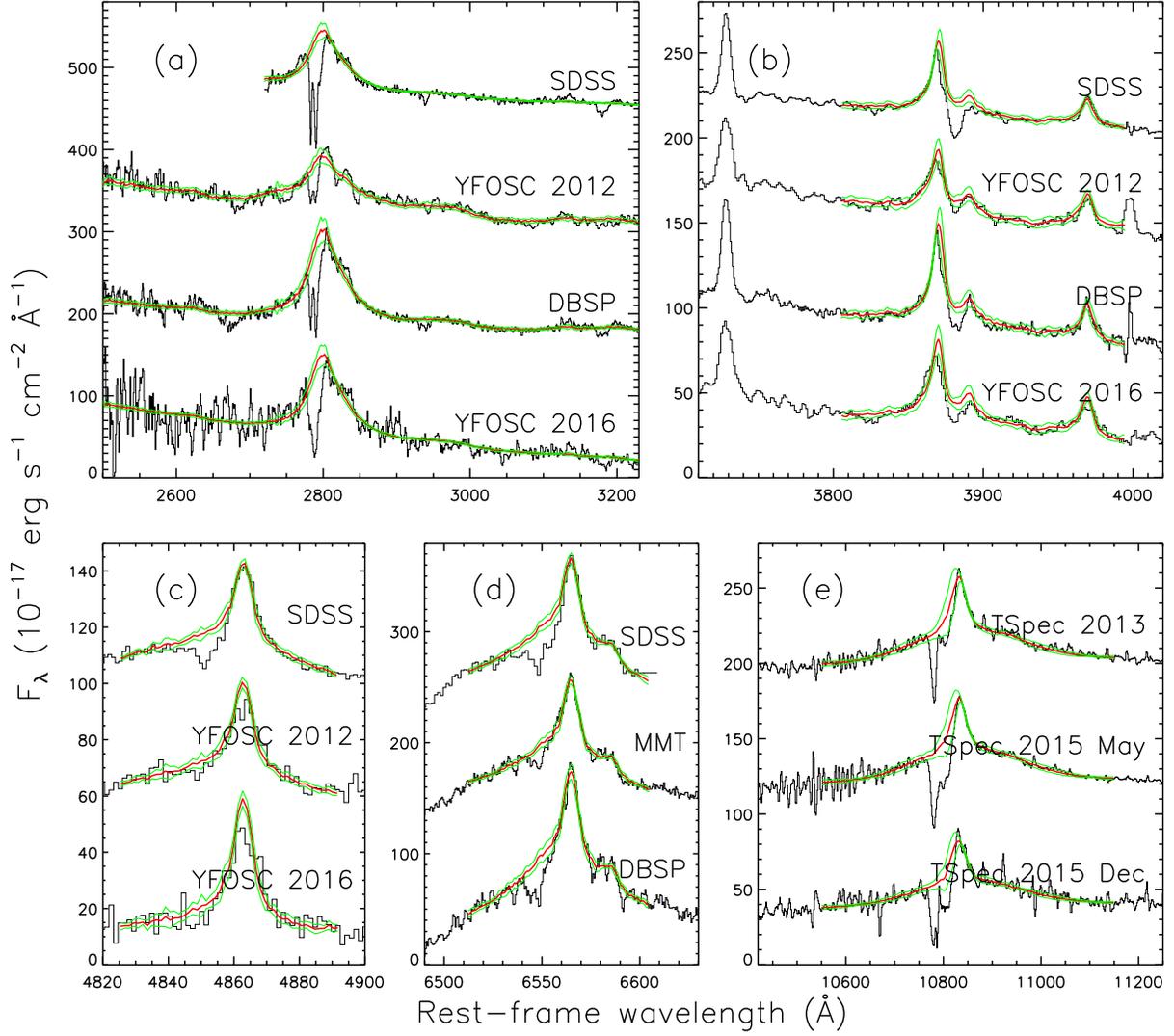}
 \caption{:
  The pair-matching results for windows of (a) Mg II and He I* $\lambda$$\lambda$2946, 3189; (b) He I* $\lambda$3889; (c) H$\beta$; (d) H$\alpha$; (e) He I* $\lambda$10830.
  The median spectra of the best 10 fittings (red lines) are plotted over the observed spectra (black lines).
  The ranges between 1-$\sigma$ standard deviation of the best 10 fittings for each data point are illuminated in green lines.
  }}
\label{pairmatch}
\end{figure}

\begin{figure}
\centering{
 \includegraphics[scale=0.98]{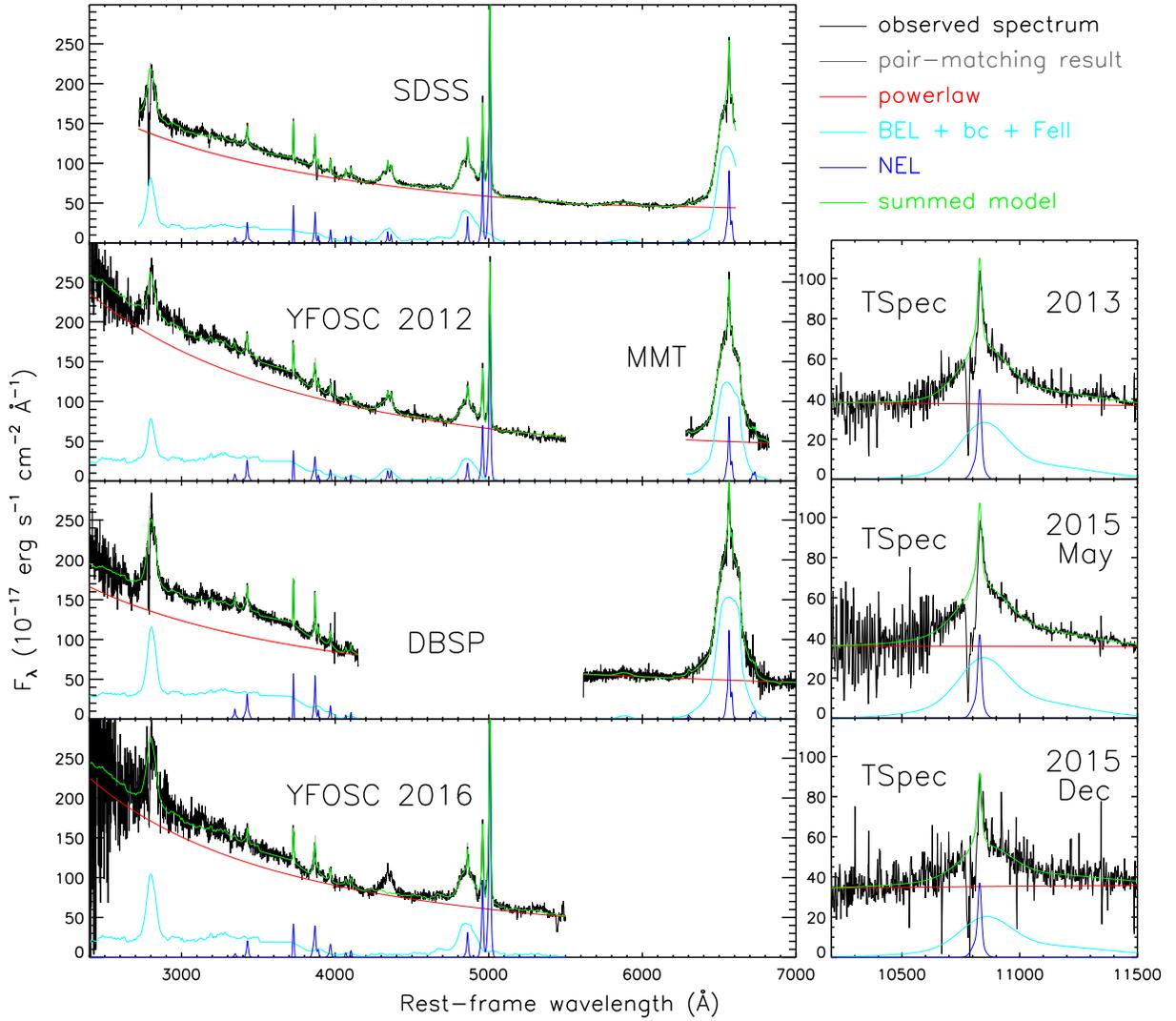}
 \caption{:
  The decomposition results of the absorption-free spectra.
  We show the observed spectra in black and the absorption-free spectra in grey, along with the best-fitting power-law (red), BELR (cyan) and NELR (blue) models, and the summed of the three (green).
  }}
\label{decomp}
\end{figure}

\begin{figure}
\centering{
 \includegraphics[scale=0.98]{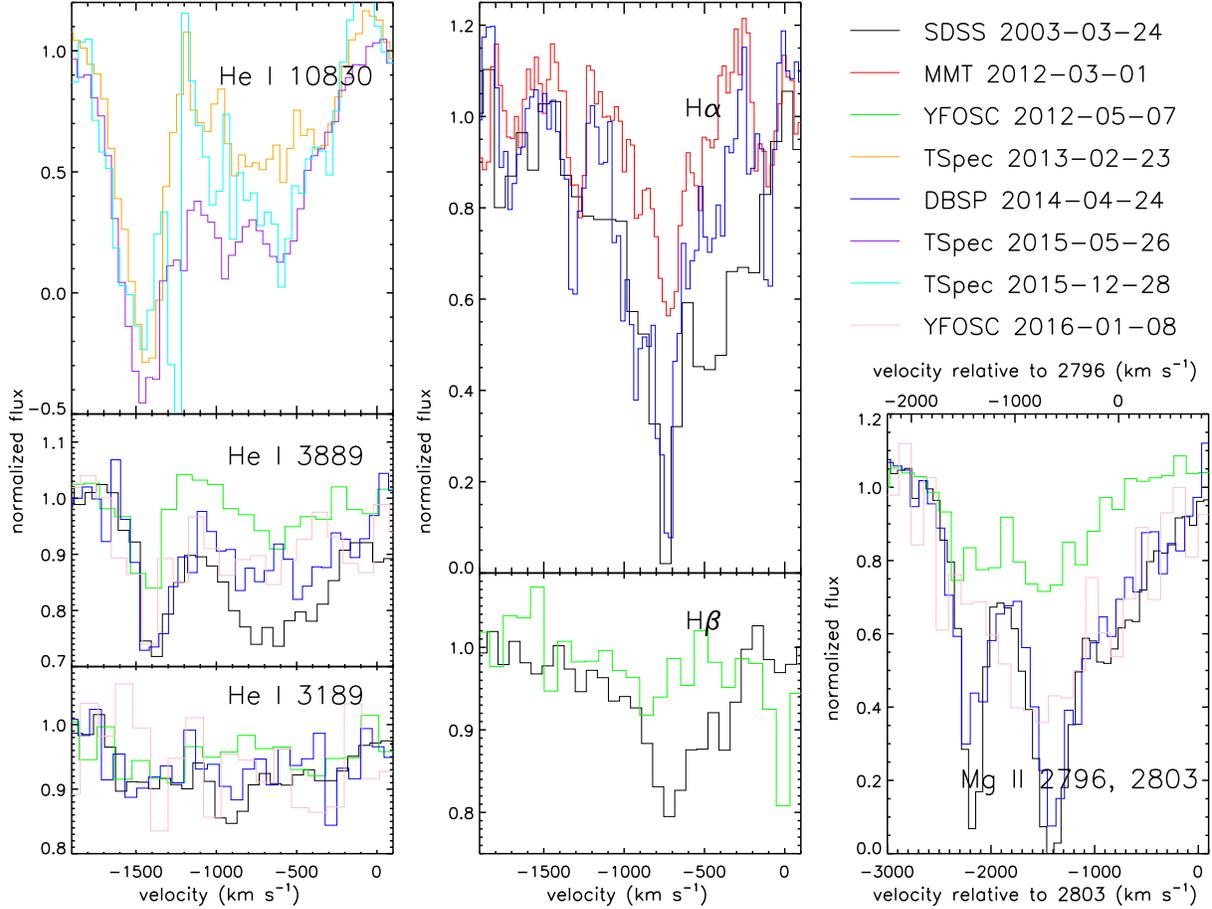}
 \caption{:
  The normalized spectra by assuming that the absorbing gas covers the whole power-law model and no BELR model.
  The absorption windows and the colors are same with Figure 3.
  For Mg II $\lambda$$\lambda$2796, 2803 doublet, we converted the wavelengths to velocities for each of the two lines.
  }}
\label{trough}
\end{figure}

\begin{figure}
\centering{
 \includegraphics[scale=0.98]{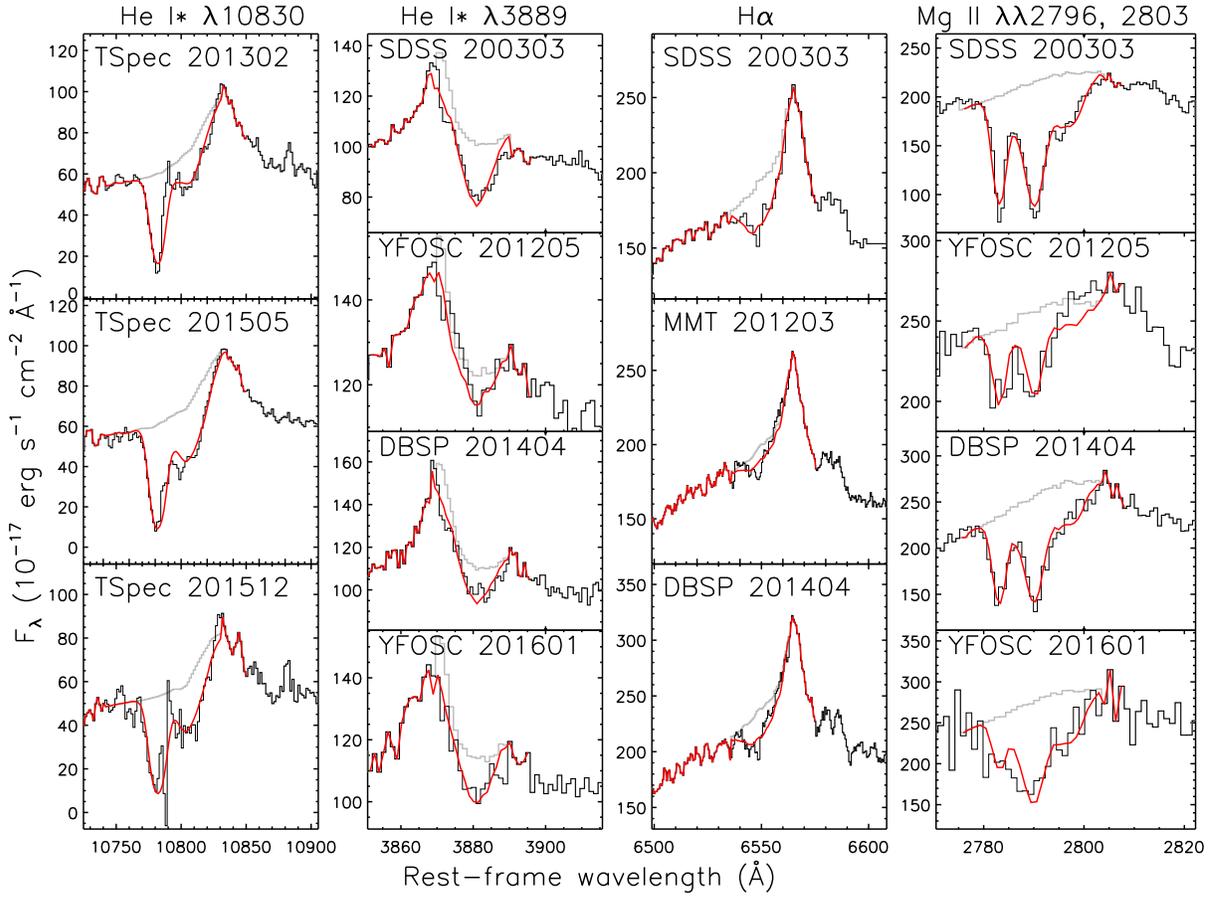}
 \caption{:
  The best-fitting absorption models (red) for He I* $\lambda$10830, He I* $\lambda$3889, H$\alpha$ and Mg II doublet, overplotted on the observed spectra (black) and the absorption-free spectra (grey).
  }}
\label{fitabs_opt}
\end{figure}

\begin{figure}
\centering{
 \includegraphics[scale=0.98]{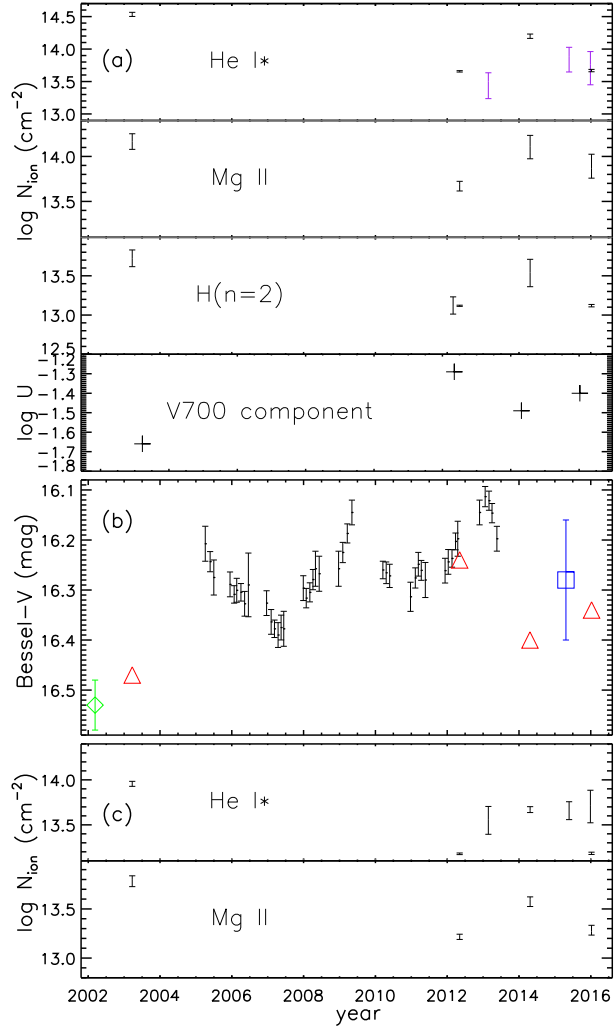}
 \caption{:
  (a) The column densities of He I*, Mg II and H(n=2) ions and ionization parameter $U$ as functions of observation time for the V700 absorbing gas.
  (b) The V-band light curve (same as Figure 2).
  (c) Same as (a), but for the V1400 absorbing gas.
  }}
\label{lc_nion}
\end{figure}

\begin{figure}
\centering{
 \includegraphics[scale=0.98]{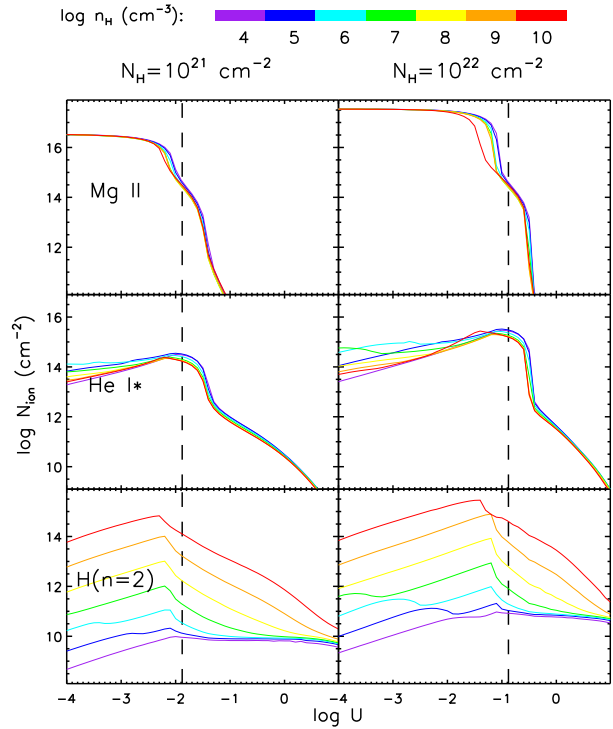}
 \caption{:
  The column density of the Mg II, He I* and H(n=2) ions as functions of ionization parameter U for electron densities from $10^4$ to $10^{10}$ cm$^{-3}$ and gas column densities of $10^{21}$ and $10^{22}$ cm$^{-2}$.
  The black dashed lines are given by $U=\frac{N_H \cdot \beta}{c}$, where $\beta$ is the H recombination coefficient for $T=10^4$ K.
  This means that the ionization front is well developed in the area on the left of the line and is not reached on the right.
  }}
\label{nioncurve}
\end{figure}

\begin{figure}
\centering{
 \includegraphics[scale=0.98]{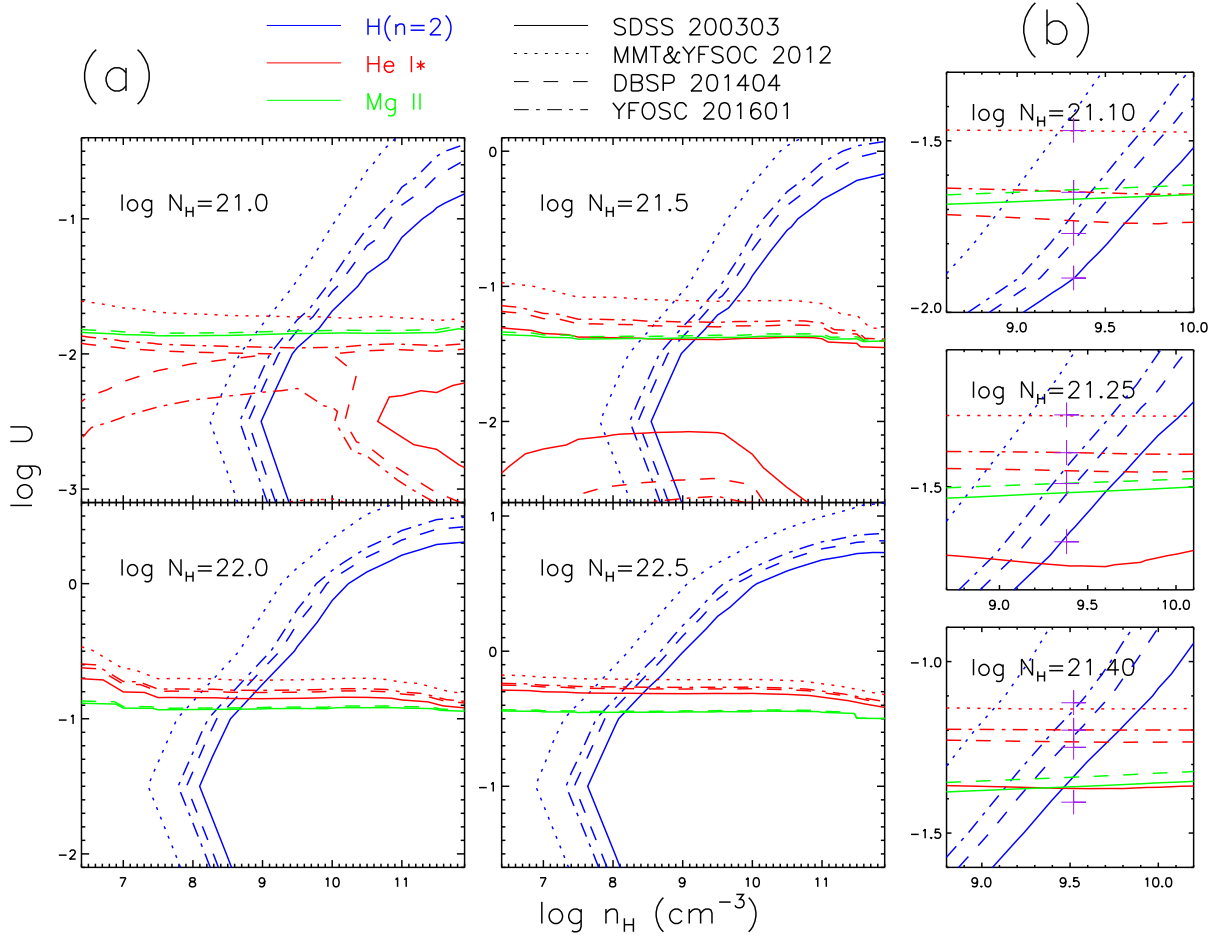}
 \caption{:
  \textbf{(a)}:
  The contours of the column densities of H(n=2) (blue), He I* (green) and Mg II (red) ions as functions of $n_H$ and $U$ for V700 absorbing components, according to the levels of observed values from SDSS 200303 (valid), YFOSC\&MMT 2012 (dotted), DBSP 201404 (dashed) and YFOSC 201601 (dot dashed) spectra.
  The $N_{\rm H(n=2)}$ value of YFOSC are obtained from MMT spectra due to smaller uncertainty.
  The contours are plotted using the sparse grid of models for log$N_H$ values from 21 to 22.5 in 0.5 dex and $N_H$ value for each is labeled.
  \textbf{(b)}:
  The same as (a), but using a denser grid of models and for $N_H$ values from 21.1, 21.25 and 21.4.
  The best-fitting solutions are labeled in purple pluses.
  }}
\label{varU}
\end{figure}

\begin{figure}
\centering{
 \includegraphics[scale=0.93]{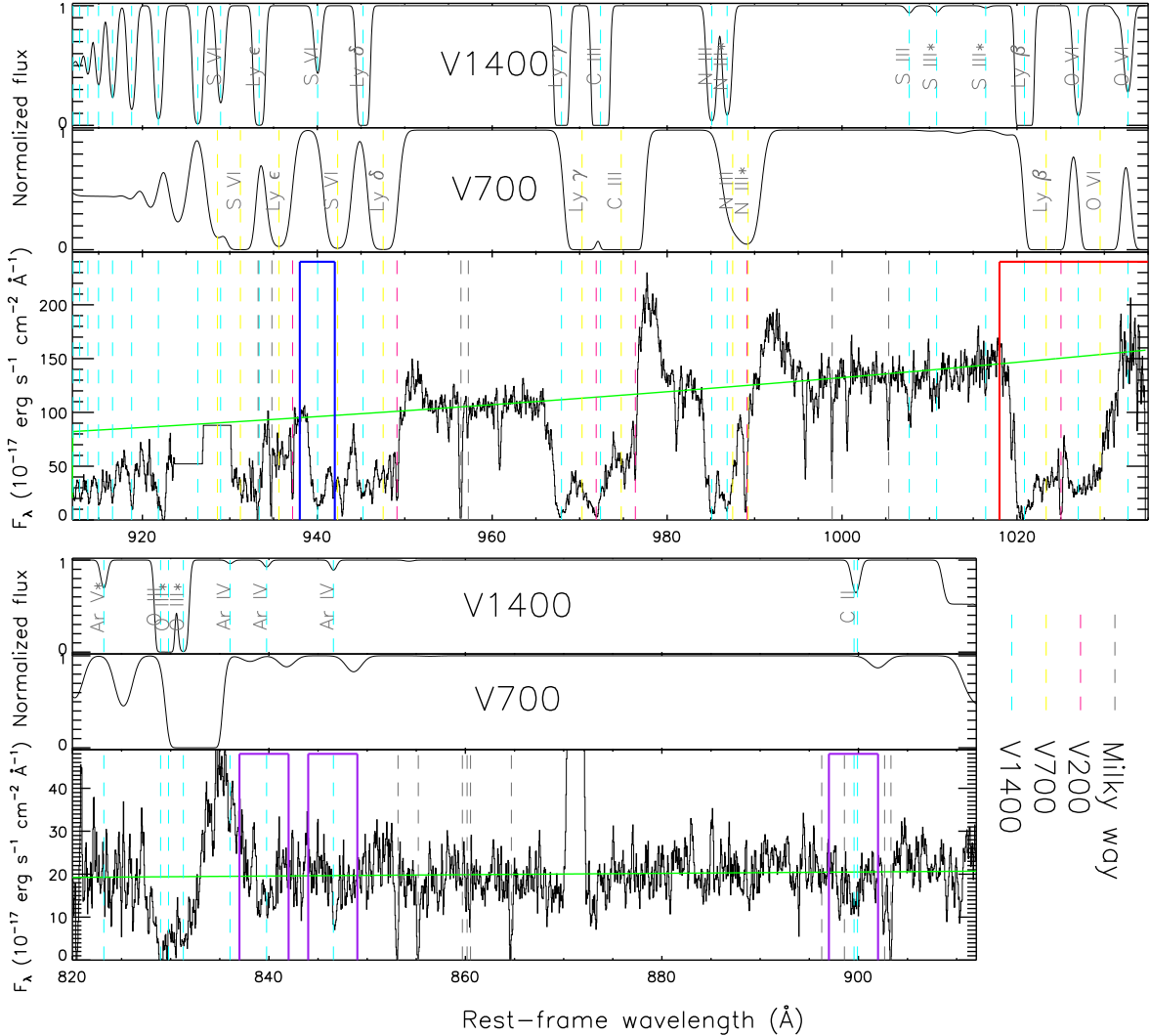}
 \caption{:
  \textbf{upper panel}:
  The predicted absorption spectra for V1400 and V700 components from Cloudy simulations and the observed spectrum in 912--1034 \AA.
  When generating absorption spectra, we only considered the elements with abundance $>10^{-6}$ relative to Hydrogen and the lines with strong enough integrated optical depths.
  For C II, O III, N III, S III and Ar V, which have excited states with low excitation energy, we calculated the populations of ground and excited states for two situations, high density limit and log$n_H=3$ for the V700 and V1400 components, respectively.
  We labeled the ion name generating the main absorption lines in the simulation.
  The dashed vertical lines represents the line centers of V1400 components (cyan), V700 component (yellow), V200 component (pink) and galactic foreground (grey).
  In the observed spectrum part, we plotted the continuum using green line.
  We also labeled regions which is noteworthy using boxes and discussed them detailedly in the main text.
  \textbf{lower panel}:
  The same as the upper panel, but for spectral range of 820--912 \AA.
  }}
\label{fakecos_naive}
\end{figure}

\begin{figure}
\centering{
 \includegraphics[scale=0.98]{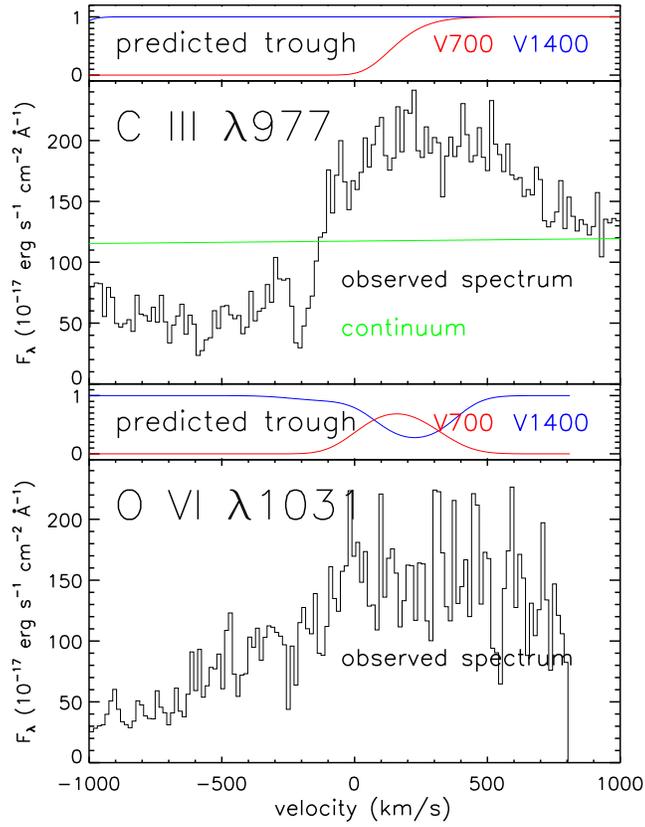}
 \caption{:
  The observed spectrum in velocity space of C III $\lambda$977 and O VI $\lambda$1031, illustrating the emission line profile.
  }}
\label{emline_fwhm}
\end{figure}

\begin{figure}
\centering{
 \includegraphics[scale=0.98]{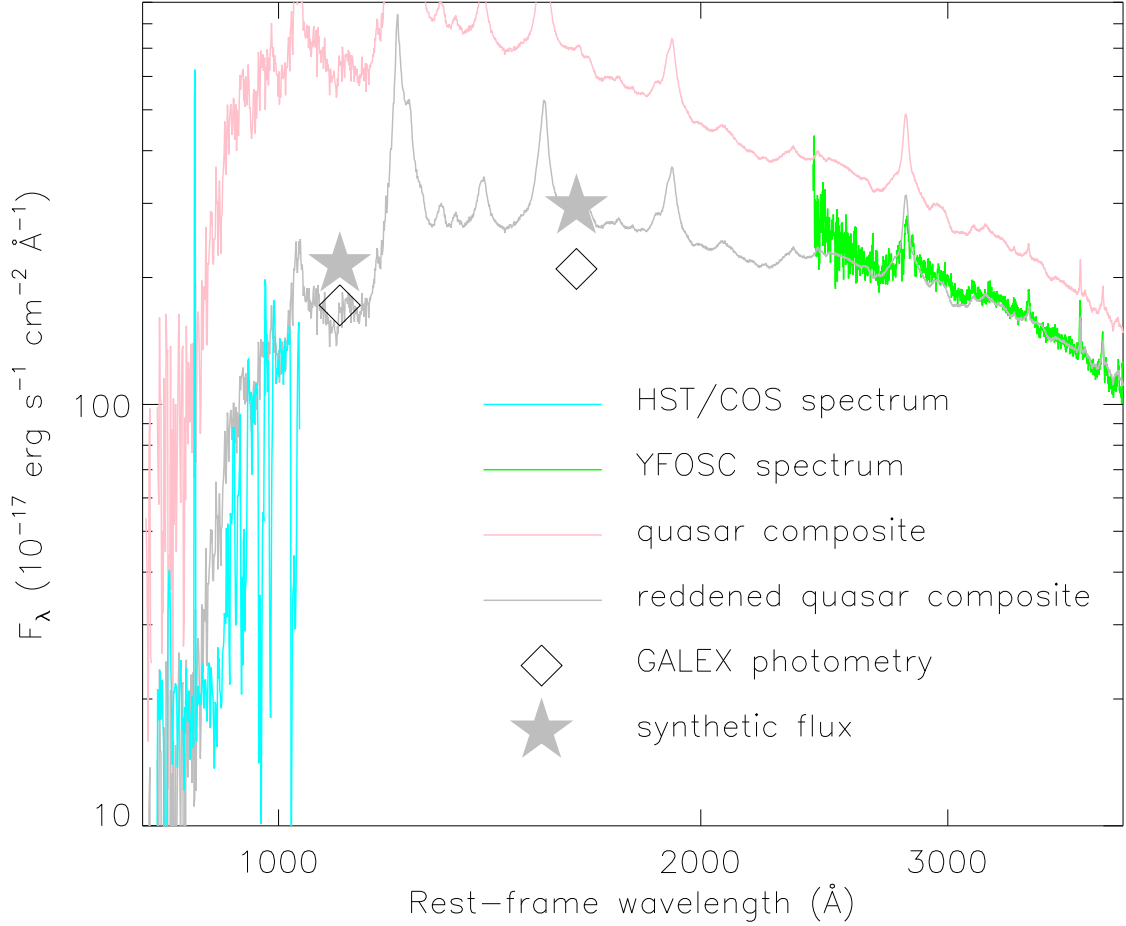}
 \caption{:
  Comparison between the observed HST/COS 2010 (cyan) and YFOSC 2012 (green) spectra and the reddened composite spectrum (grey) with $E_{\rm B-V}=0.07$.
  We also plotted the unreddened quasar composite spectrum for the illustration of the dust extinction of the HST/COS spectrum.
  The black diamonds show the GALEX photometry in FUV and NUV band, and the grey stars show the synthetic flux using the reddend composite spectrum in these two band.
  }}
\label{uvsed}
\end{figure}

\begin{figure}
\centering{
 \includegraphics[scale=0.98]{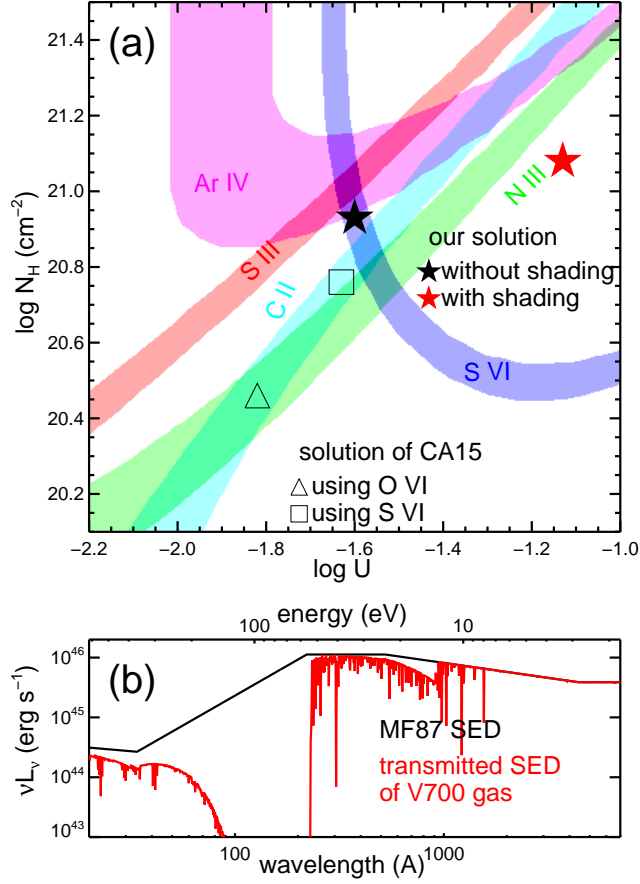}
 \caption{:
  \textbf{(a)}:
  The contours of column densities of C II, N III, S III, S VI and Ar IV as functions of $N_H$ and $U$ for V1400 absorbing gas.
  The $n_H$ value is fixed to be $10^3$ cm$^{-3}$ in the Cloudy simulation.
  The best-fitting solutions by considering and not considering the shading effect are labeled using red and black stars.
  We also labeled the solution from CA15 with triangle and the solution using S VI from CA15 simulation with square.
  \textbf{(b)}:
  Comparison of the MF87 SED (black) and transmitted SED of V700 absorbing gas (red).
  }}
\label{cloudy_highv}
\end{figure}

\begin{figure}
\centering{
 \includegraphics[scale=0.98]{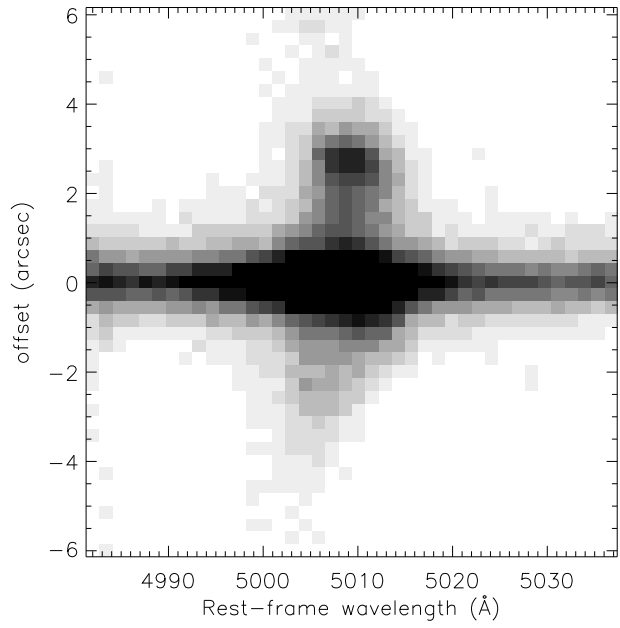}
 \caption{:
  The two dimensional spectral image from YFOSC 201205 observation in [O III] $\lambda$5007 region, illustrating the EELR.
  }}
\label{eelr}
\end{figure}

\clearpage

\begin{appendix}
\section{The details of the pair-matching method}

In the main body, we introduced the principle of the pair-matching method.
Here we describe in detail the application of this method for LBQS 1206+1052.
We divided four ranges containing absorption lines as follows:

The first range is 2500--3300 \AA\ for Mg II doublet and He I $\lambda\lambda$2946, 3189 lines.
Liu15 developed the pair-matching method for this range.
We directly used the template library of unabsorbed quasar of Liu15, which contains 1343 quasars from SDSS DR7 quasar catalog.
We noticed that the EW of Mg II emission line varies among the observed spectra of LBQS 1206+1052.
Thus we made a modification to the method of Liu15.
In brief, we decomposed the contributions of pseudo-continuum and Mg II emission lines for the templates following Wang et al. (2009), and applied different scaling factors to the two components when fitting different spectra of LBQS 1206+1052.
This can be expressed as:

\begin{equation}
f_{obs,i}(\lambda) = f_{t,{\rm conti}}(\lambda) \times S_{1,i} + f_{t,{\rm MgII}}(\lambda) \times S_{2,i} + (P_{0,i} + P_{1,i} \lambda + P_{2,i} \lambda^2),
\end{equation}

where the subscript i indicate different spectra for LBQS 1206+1052; $f_{t,{\rm conti}}$ and $f_{t,{\rm MgII}}$ are the pseudo-continuum and Mg II components of the templates; $S_{1,i}$ and $S_{2,i}$ are the scaling factors for the two components, and $P_{0,i}$, $P_{1,i}$, $P_{2,i}$ are coefficients for an additional two-order polynomial.
These five parameters are variable for different spectra.
Thus the variations in strength of BEL, strength of Fe II emission, strength and spectral shape of continuum are all considered.
We masked three ranges of 2775--2804, 2930--2945 and 3170--3190 \AA\ in the fitting which are affected by absorption lines.
We noticed that the spectral resolutions in Mg II region, which are listed in Table 1, are not uniform.
Thus we blurred the templates using Gaussians with different widths for different spectra to be accounted for resolutions.
We also considered the effect of resolutions for the following two parts in optical.

The second range is 3800--4000 \AA\ for He I $\lambda$3889 lines.
Liu15 also developed the pair-matching method for this range.
However, the situation in LBQS 1206+1052 is different as the absorption trough is strongly affected by [Ne III] $\lambda$3869 and H$\zeta$ emission lines.
Thus we further developed the pair-matching method as follows.
We reselected the unabsorbed templates from DR7 and DR12 quasar catalogs.
We required $z<1.2$ for DR7 and $z<1.3$ for DR12 to ensure that He I* $\lambda$3889 locates in the spectral coverage, and required that the median SNR $>10$ in 3800--4000 \AA.
We found variation in EW of NEL among spectra of LBQS 1206+1052, thus we decomposed the templates spectra into two components, a NEL component and a continuum+BEL component.
In brief, we considered Balmer BELs from H$\delta$ to H$\zeta$, and NELs of [O II] $\lambda$3727, [Ne III] $\lambda\lambda$3869,3968, H$\zeta$, H$\epsilon$ and H$\delta$, and assumed that the continuum is a two-order polynomial and that all of BELs and NELs are Gaussian, and fit the template spectra in 3700--4200 \AA.
To better recover the shape of [Ne III] $\lambda$3870 emission lines, we only selected those with SNR of [Ne III] $>10$.
We also visually ruled out those with absorption lines and finally selected a library of 576 quasar spectra as templates.
We fit the spectra of LBQS 1206+1052 using this formula:

\begin{equation}
f_{obs,i}(\lambda) = f_{t,{\rm conti+BEL}}(\lambda) \times S_{1,i} + f_{t,{\rm NEL}}(\lambda) \times S_{2,i} + (P_{0,i} + P_{1,i} \lambda + P_{2,i} \lambda^2),
\end{equation}

which is similar to that used for Mg II region.
In the fitting we raised the weight in a region of 3840--3869 \AA, where the blue side of [Ne III] $\lambda$3869 NEL is in, because it is the key region to recover the whole shape of the NEL.
The region of 3869--3890 \AA, corresponding to $-1500$--0 \kmsb for He I* $\lambda$3889, was masked in the fitting, and the velocity range was determined using the He I* $\lambda$10830 trough.

The third range is for H$\beta$ and H$\alpha$ absorption troughs.
The unabsorbed templates are also selected from DR7 and DR12 quasar catalogs, and $z<0.4$ was required for DR7 and $z<0.5$ for DR12 to ensure both of the two lines lie in the spectral coverage.
To recover the absorption-free spectra, the main features that we need to consider are H$\alpha$, H$\beta$ and [N II] $\lambda$6548 NELs, and H$\alpha$ and H$\beta$ BELs.
Therefore we selected a large range containing these features at first.
The two parts are treated as a whole because the profiles of H$\alpha$ and H$\beta$ emission lines are strongly correlated.
After a test, we found there are too few templates that can fit the entire emission line profile consisting NELs and BELs at the same time.
Thus we narrowed down the wavelength range to 4825--4892 and 6510--6610 \AA\ in pair-matching process because only the top part of the BEL profile is essential for recovering the absorption-free spectrum.
We required that the median SNR in the two ranges are both $>$25, and after this 1496 unabsorbed quasars remain.
The variation in continuum and BEL can be expressed approximately as a two-order polynomial for H$\beta$ and H$\alpha$ part each.
Thus we have:

\begin{equation}
\left\{
\begin{array}{lr}
 f_{obs,i}(\lambda) = f_t(\lambda) \times S_{1,i} + (P_{0,i} + P_{1,i} \lambda + P_{2,i} \lambda^2), \ \ \ \ 4825<\lambda<4892, \\
 f_{obs,i}(\lambda) = f_t(\lambda) \times S_{2,i} + (Q_{0,i} + Q_{1,i} \lambda + Q_{2,i} \lambda^2), \ \ \ \ 6510<\lambda<6610
\end{array}
\right.
\end{equation}

We masked the regions affected by absorption lines using a velocity range of $-1100<v<-100$ for H$\beta$ and H$\alpha$ each, and the velocity range was set according to the absorption parameter from Ji12.

The last range was for He I* $\lambda$10830 trough.
The wavelength range was selected to be 10550--11150 \AA, which contains emission-line-free regions for both blue and red sides.
We collected NIR quasar spectra from Glikman et al. (2006), Riffel et al. (2006) and Landt et al. (2008), and added some quasar spectra we observed in the past using TripleSpec.
We selected unabsorbed templates by three criteria as follows.
First, the He I* $\lambda$10830 does not exceed the wavelength coverage, or fall in the gap between J and H or in the gap between H and K.
Second, a BEL can be clearly seen, which means that the spectrum does not show only He I* $\lambda$10830 NEL (FWHM$<$2000 \kmsb), and that the emission line has SNR $>$15.
Third, there are no He I* $\lambda$10830 absorption lines by visual check.
After the selection, 51 quasar spectra remained and were used as templates.
We also decomposed the spectra templates considering the variation in continuum and BEL.
The model to fit the templates consists of a two-order polynomial as continuum, and two Gaussians for He I* $\lambda$10830 and Pa$\gamma$ BELs, and two Gaussians for the two NELs.
And the spectra of LBQS 1206+1052 were fit using:

\begin{equation}
f_{obs,i}(\lambda) = f_{t,{\rm conti}}(\lambda) \times S_{1,i} + f_{t,{\rm BEL}}(\lambda) \times S_{2,i} + f_{t,{\rm NEL}}(\lambda) \times S_{3,i} + (P_{0,i} + P_{1,i} \lambda),
\end{equation}

while we masked a wavelength range of 10741--10831 \AA, which was determined visually.
For He I* $\lambda$10830, we only used a one-order polynomial for the deviation in continuum because we found that a one-order polynomial is enough.

\section{Testing the fitting procedure}

{ We examined the robustness of the fitting procedure to obtain the BAL parameters and their uncertainties.
We made the test in Mg II region.
We randomly selected 100 unabsorbed quasars spectra from the library with total 1343 quasars.
We generated fake absorbed spectra as:
\begin{equation}
f_{\rm fake}(\lambda) = f_{\rm unabsorbed}(\lambda) \times e^{-\tau(\lambda)}
\end{equation}
where $\tau(\lambda)$ is the same for all the 100 spectra, and has a Gaussian profile with three parameters centroid $v=-1400$ \kms, $\sigma=250$ \kmsb and integrated optical depth of Mg II $\lambda$2803 $\tau_{2803}=4$.
We fit the fake absorbed spectra using the method described in Section 3.1.
The results are plotted in Figure 16.
For all the three parameters, the best values are close to the ture value.
For 92 of the 100 unabsorbed quasar, the true value of centroid falls into the 90\% confidence interval, and the corresponding numbers for $\sigma$ and $\tau_{2803}$ are 88 and 84, respectively.
The mean fraction of 88\% is close to 90\%, thus the method yields reliable confidence interval.

We also examined our method for directly measuring BAL EW.
We also measured from the 100 fake absorbed spectra, and plotted the results in Figure 16.
The best values are close to the true value of 4.89 \AA, and for 88 of 100 the true value is in the 90\% confidence interval.
Thus the method is also robust when measuring BAL EW.}

\clearpage

\begin{figure}
\centering{
 \includegraphics[scale=0.98]{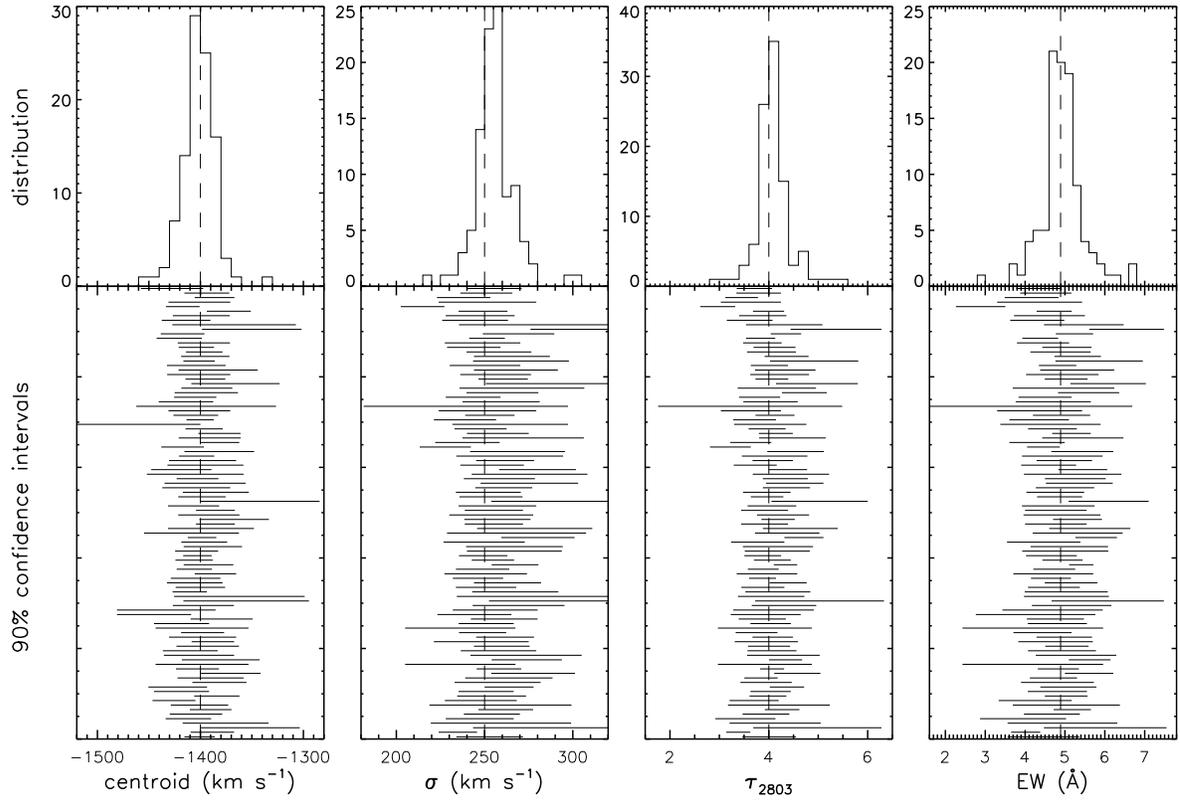}
 \caption{:
  Results of testing our method to fit BAL parameters or to measure EW.
  From left to right are the results of centroid, $\sigma$, $\tau_{2803}$ and EW.
  The upper panel shows the distribution of the best values for the 100 fake absorbed spectra, and the lower panel shows the 90\% confidence interval.
  }}
\label{testpairmatch}
\end{figure}

\end{appendix}

\end{document}